\documentclass{aa}  

\usepackage{graphicx}
\usepackage{txfonts}
\usepackage{subcaption}
\usepackage{natbib}
\usepackage{multirow}
\usepackage{bigstrut}
\usepackage{longtable,threeparttablex,booktabs,needspace}
\usepackage{lscape}
\usepackage{epigraph}
\usepackage{multirow,bigdelim}
\usepackage{etoolbox}

\bibpunct{(}{)}{;}{a}{}{,} 

\newcommand{\propanal}{C$_2$H$_5$CHO}
\newcommand{\ethyleneoxide}{c-C$_2$H$_4$O}
\newcommand{\acetone}{CH$_3$COCH$_3$}

\newcommand{\acetaldehyde}{CH$_3$CHO}

\begin{document}

 \title{The ALMA-PILS survey: First detections of ethylene oxide, acetone and propanal toward the low-mass protostar IRAS~16293-2422}
 \titlerunning{The ALMA-PILS survey: First detections of ethylene oxide, acetone and propanal toward IRAS 16293-2422}

   \author{J. M. Lykke\inst{1} \and A. Coutens\inst{2} \and J.~K. J{\o}rgensen\inst{1} \and M.~H.~D. van der Wiel\inst{1} \and R.~T. Garrod\inst{3} \and H.~S.~P. M\"{u}ller\inst{4} \and P. Bjerkeli\inst{1,5} \and T.~L.~Bourke\inst{6} \and H.~Calcutt\inst{1} \and M.~N. Drozdovskaya\inst{7} \and C.~Favre\inst{8,9} \and E.~C. Fayolle\inst{10} \and S.~K.~Jacobsen\inst{1} \and K.~I.~\"{O}berg\inst{10} \and M.~V.~Persson\inst{7} \and E.~F.~van Dishoeck\inst{7,11} \and S.~F.~Wampfler\inst{12}
          }
    \authorrunning{Lykke et al. }

   \institute{Centre
  for Star and Planet Formation, Niels Bohr Institute \& Natural
  History Museum of Denmark, University of
  Copenhagen, {\O}ster Voldgade 5--7, DK-1350 Copenhagen {K}., Denmark
  \and Department of Physics and Astronomy, University College London, Gower St., London, WC1E 6BT, UK
  \and Departments of Chemistry and Astronomy, University of Virginia, Charlottesville, VA 22904, USA
  \and I. Physikalisches Institut, Universit\"{a}t zu K\"{o}ln, Z\"{u}lpicher Str. 77, 50937 K\"{o}ln, Germany
  \and Department of Earth and Space Sciences, Chalmers University of Technology, Onsala Space Observatory, 439 92 Onsala, Sweden
  \and SKA Organization, Jodrell Bank Observatory, Lower Withington, Macclesfield, Cheshire SK11 9DL, UK
  \and Leiden Observatory, Leiden University, PO Box 9513, NL-2300 RA Leiden, The Netherlands
  \and Universit\'e Grenoble Alpes, IPAG, F-38000 Grenoble, France
  \and CNRS, IPAG, F-38000 Grenoble, France
  \and Harvard-Smithsonian Center for Astrophysics, 60 Garden Street, Cambridge, MA 02138, USA
  \and Max-Planck Institut f\"{u}r Extraterrestrische Physik (MPE), Giessenbachstr. 1, 85748 Garching, Germany
  \and Center for Space and Habitability (CSH), University of Bern, Sidlerstrasse 5, CH-3012 Bern, Switzerland
}             

   \date{}

\offprints{J.~M.~Lykke, \email{juliemarialykke@gmail.com}}

 
  \abstract
   {One of the open questions in astrochemistry is how complex organic and prebiotic molecules are formed. The unsurpassed sensitivity of the Atacama Large Millimeter/submillimeter Array (ALMA) takes the quest for discovering molecules in the warm and dense gas surrounding young stars to the next level.}
   {Our aim is to start the process of compiling an inventory of oxygen-bearing complex organic molecules toward the solar-type Class 0 protostellar binary IRAS~16293-2422 from an unbiased spectral survey with ALMA, Protostellar Interferometric Line Survey (PILS). Here we focus on the new detections of ethylene oxide (\ethyleneoxide{}), acetone (\acetone{}), and propanal (\propanal{}).}
   {With ALMA, we surveyed the spectral range from 329 to 363~GHz at 0.5$''$ (60~AU diameter) resolution. Using a simple model for the molecular emission in local thermodynamical equilibrium, the excitation temperatures and column densities of each species were constrained.}
   {We successfully detect propanal (44 lines), ethylene oxide (20 lines) and acetone (186 lines) toward one component of the protostellar binary, IRAS16293B. The high resolution maps demonstrate that the emission for all investigated species originates from the compact central region close to the protostar. This, along with a derived common excitation temperature of $T_\mathrm{ex} \approx$ 125~K, is consistent with a coexistence of these molecules in the same gas.}
   {The observations mark the first detections of acetone, propanal and ethylene oxide toward a low-mass protostar. The relative abundance ratios of the two sets of isomers, a \acetone{}/\propanal{} ratio of 8 and a \acetaldehyde{}/\ethyleneoxide{} ratio of 12, are comparable to previous observations toward high-mass protostars. The majority of observed abundance ratios from these results as well as those measured toward high-mass protostars are up to an order of magnitude above the predictions from chemical models. This may reflect either missing reactions or uncertain rates in the chemical networks. The physical conditions, such as temperatures or densities, used in the models, may not be applicable to solar-type protostars either.
}


\maketitle
%
\section{Introduction}
An important task of modern-day astrochemistry is to understand how complex organics and possible pre-biotic molecules form near young stars. The high sensitivity and angular and spectral resolution of the Atacama Large Millimeter/submillimeter Array (ALMA) enables detection of molecular species with faint emission lines in otherwise confused regions. The capabilities of ALMA were demonstrated early on by the first detection of the prebiotic molecule glycolaldehyde toward the low-mass protostar, IRAS~16293-2422 \citep{Jorgensen2012}. This detection illustrates the potential for imaging emission from the simplest building blocks for biologically relevant molecules during the earliest stages of the Solar System on the scales where protoplanetary disks emerge, and for understanding how these molecules are formed and in what abundances. This paper presents the first detections of three such species, ethylene oxide (\ethyleneoxide{}), propanal (\propanal{}) and acetone (\acetone{}) toward IRAS~16293-2422 from an unbiased spectral survey with ALMA (Protostellar Interferometric Line Survey or PILS; \citealt{Jorgensen2016}).

Traditionally, detections of complex organic molecules have mostly been associated with the hot cores around high-mass protostars toward the warm and dense central regions around such luminuous sources where the molecules sublimate from the icy mantles of dust grains. Some low-mass protostars show similar characteristics on small scales; the so-called \textit{hot corinos} \citep{vanDishoeckandBlake1998,Bottinelli2004,Ceccarelli2004}. A prime example of this is IRAS~16293-2422 (IRAS16293 hereafter), a protostellar Class 0 binary system, located at a distance of 120~pc \citep{Loinard2008}. IRAS16293 is perhaps the best low-mass protostellar testbed for astrochemical studies \citep[see, e.g.,][]{Blake1994,vanDishoeck1995,Ceccarelli2000,Schoier2002}. It has the brightest lines by far of all well-studied low-mass protostars and shows detections of a wealth of complex organic molecules \citep{Cazaux2003,Caux2011}. These complex organics arise in the dense gas around each of its two binary components that each show distinct chemical signatures in the warm gas on small scales resolved by (sub)millimeter wavelength aperture synthesis observations \citep{Bottinelli2004,Kuan2004,Bisschop2008,Jorgensen2011}.

To understand how these complex organic molecules form, combinations of systematic studies establishing large inventories of similar organic molecules are needed. For this purpose, structural isomers are particularly interesting since they usually share some formation and destruction pathways. The relative abundance of two such isomers may therefore provide important constraints on astrochemical models. Examples of such interesting isotope pairs are ethylene oxide and acetaldehyde as well as acetone and propanal. Ethylene oxide was first detected toward the galactic center source Sagittarius B2(N) (Sgr B2(N)) by \citeauthor{Dickens1997} (\citeyear{Dickens1997}; confirmed by \citealt{Belloche2013}), and has since been observed in several massive star-forming regions \citep{Nummelin1998, Ikeda2001} but so far not toward any low-mass protostar. 
 Acetone (\acetone{}), also called propanone, was the first molecule with ten atoms to be observed in the ISM. The molecule was first detected in the hot molecular core Sgr B2 \citep{Combes1987,Snyder2002} and later in the Orion-KL star-forming region \citep{Friedel2005, FriedelandSnyder2008, Peng2013}. It was also detected toward other massive star-forming regions \citep{Isokoski2013} as well as toward an intermediate-mass protostar \citep{Fuente2014}. Several lines of the SMA survey of IRAS16293 were also assigned to acetone by \citet{Jorgensen2011}, but it has never been properly identified in this source.
 More recently, it was found in material from the comet 67P/Churyumov-Gerasimenko by the COmetary Sampling And Composition (COSAC) experiment on Rosetta's lander Philae \citep{Goesmann2015}. 
Propanal (\propanal{}) has previously been detected in Sgr B2(N) by \cite{Hollis2004b}, where it coexists with propynal and propenal. It was also detected towards two Galactic center molecular clouds by \citet{Requena2008}. Like acetone, propanal was found to be present in the comet 67P/Churyumov-Gerasimenko \citep{Goesmann2015}. 

This paper presents detections of ethylene oxide, acetone and propanal toward IRAS16293 utilising a large ALMA survey at (sub)millimeter wavelength. These are all first time detections in IRAS16293 and in low-mass protostars in general. In Sect.~\ref{sect_obs}, we briefly describe the observations. The identification and analysis of the data are presented in Sect.~\ref{sect_results}. Finally, we discuss the results in Sect.~\ref{sect_discussion} and conclude in Sect.~\ref{sect_conclu}. 

\section{Observations}
\label{sect_obs}
IRAS16293 was observed as part of the PILS program (PI: Jes K. J{\o}rgensen): the survey consists of an unbiased spectral survey covering a significant part of ALMA's Band~7 (wavelengths of approximately 0.8~mm) as well as selected windows in ALMA's Bands~3 (at approximately 100 GHz; 3~mm) and 6 ( at approximately 230~GHz; 1.3~mm). In this paper we only utilise data from the Band~7 part of the survey (project-id: 2013.1.00278.S). An observing log, a description of the data reduction and a first overview of the data are presented in \cite{Jorgensen2016} and here we only summarize a number of the key features of the Band~7 observations.

The Band~7 part of the survey covers the frequency range from 329.15~GHz to 362.90~GHz in full. Data were obtained from both the array of 12~m dishes (typically 35--40~antenna in the array at the time of observations) and the Atacama Compact Array (ACA), or ``Morita Array'', of 7~m dishes (typically 9--10~antenna in use). The pointing center was in both cases set to be a location in-between the two components of the binary system at $\alpha_{\rm J2000}=16^\mathrm{h} $32$^\mathrm{m}$22.72$^\mathrm{s}$; $\delta_{\rm J2000}=-$24\degr28\arcmin34\farcs3. In total 18~spectral settings were observed: each setting covers a bandwidth of 1875~MHz (over four different spectral windows of 468.75~MHz wide). To limit the data-rate, the data were downsampled by a factor two to the native spectral resolution of the ALMA correlator, resulting in a spectral resolution of 0.244~MHz ($\approx 0.2$~km~s$^{-1}$) over 1920 channels for each spectral window. Each setting was observed with approximately 13~minutes integration on source (execution blocks of approximately 40~minutes including calibrations) for the 12~m array and double that for the ACA.

The data for each setting were calibrated and a first imaging of the continuum was performed. Thereafter, a phase-only self-calibration was performed on the continuum images and applied to the full datacubes before combining the 12~m array and ACA data and performing the final cleaning and imaging. The resulting spectral line datacubes have an root mean square (RMS) noise for the combined datasets of approximately 6--8~mJy~beam$^{-1}$~channel$^{-1}$, which translates into a uniform sensitivity better than 5~mJy~beam$^{-1}$~km~s$^{-1}$ with beam sizes ranging from $\approx$0.4--0.7$''$ depending on the exact configuration at the date of observation. The data used in this paper were produced with a circular restoring beam of 0.5$\arcsec$ to facilitate the analysis across the different spectral windows. The conversion from Rayleigh-Jeans temperature T$_\mathrm{b}[\mathrm{K}]$ to flux density $S_{\nu} [\mathrm{Jy/beam}] $ follows the standard formulation and T$_\mathrm{b}$/$S_{\nu}$ ranges from 37.2 to 45.2 K\,Jy$^{-1}$ depending on the frequency. The resulting image cubes are strongly line-confused toward the locations of the two primary protostars. A subtraction of the continuum was therefore done statistically for each spectral window (for continuum maps and more details see \citealt{Jorgensen2016}). The continuum baseline for each window is found to be robust to within twice the RMS in each channel. 

\section{Analysis and results}
\label{sect_results}
\begin{figure}[!t]
\resizebox{\hsize}{!}{\includegraphics{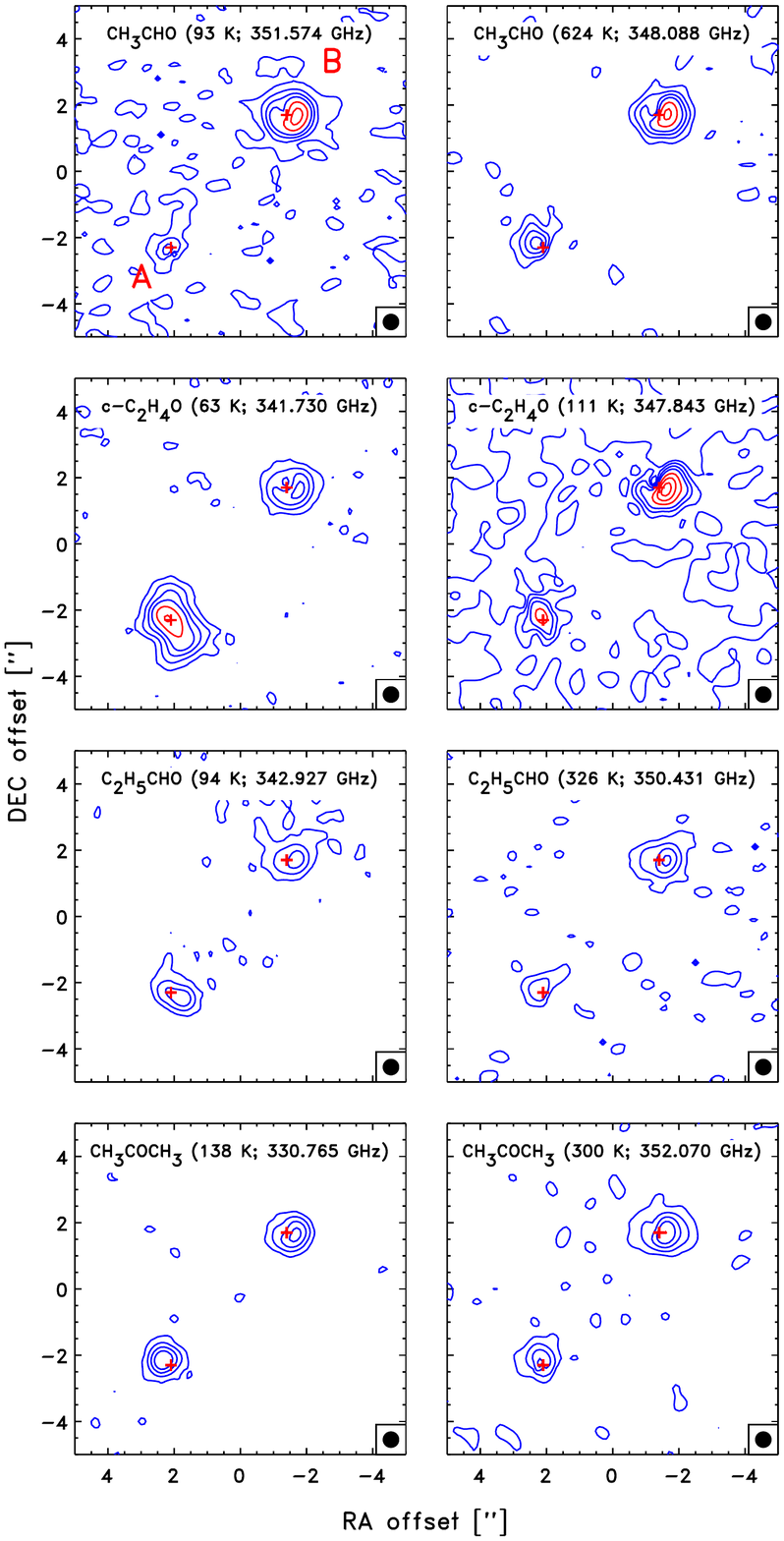}}
\caption{Integrated intensity maps of the line emission for acetaldehyde, ethylene oxide, acetone, and propanal. Left and right columns show maps for transitions with lower and higher $E_\mathrm{up}$, respectively. The locations of IRAS16293A (southeast) and IRAS16293B (northwest) are marked by the red plus-signs. The blue contours represent 4, 8, 12 and 16 $\sigma$ while the red contours show 24, 30, 36 $\sigma$, where $\sigma$ is 5 mJy beam$^{-1}$ km s$^{-1}$ for the integrated intensity. A representative beam of 0.5$''$ is shown in the lower right-hand corner of each panel.}
\label{Fig:maps}
\end{figure}

\begin{figure}[!t]
\centering
\resizebox{\hsize}{!}{\includegraphics{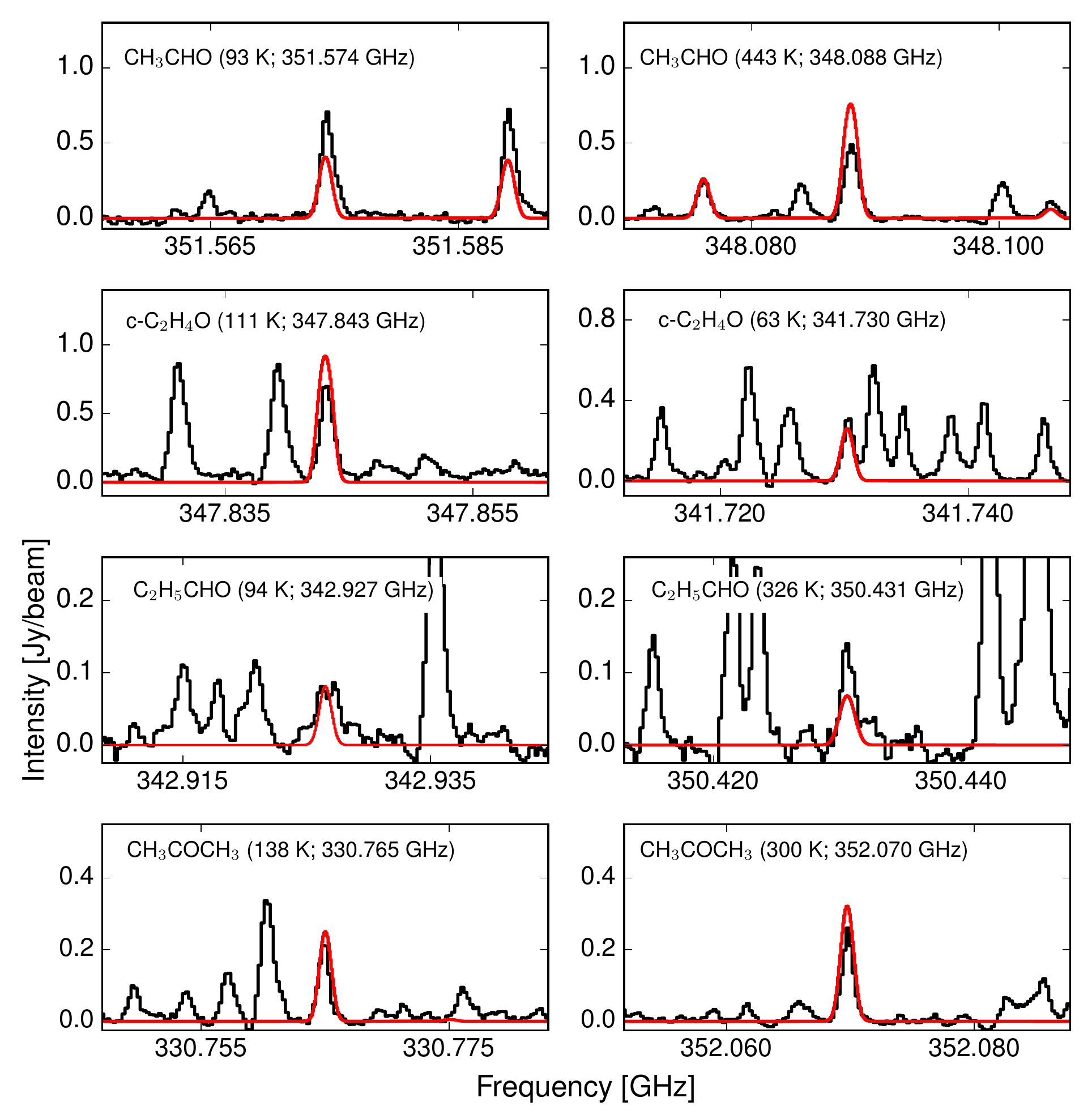}}
\caption{Observed and synthetic spectra of the representative transitions shown in Fig.~\ref{Fig:maps}. The observed spectra are extracted at a position ($-$0.45$\arcsec$ ; $-$0.30$\arcsec$) southwest of the continuum peak of IRAS16293B.}
\label{Fig:lines}
\end{figure}

Interferometric emission maps of two representive lines each for propanal, acetone, ethylene oxide, and acetaldehyde are shown in Fig.~\ref{Fig:maps}. The maps show emission toward both protostellar sources. Generally the lines toward IRAS16293A are approximately a factor five broader than toward IRAS16293B \citep[e.g.,][]{Bottinelli2004,Jorgensen2011}, which makes identification of individual species challenging. Consequently IRAS16293B is therefore better for separation of blended lines and identification of new species and in this paper we focus on that source. A comparison of the maps for the different molecules shows that the emission is marginally resolved toward IRAS16293B, consistent with a deconvolved extent of $\approx 0.5 \arcsec$ toward the location of the protostar for all species. We can therefore assume that these particular molecules coexist and trace the same gas. Extracting a spectrum from the pixel located on the peak position will give the highest emission signal, but since the continuum is optically thick and very bright there are also very prominent absorption lines in the spectrum. To reduce the influence of absorption while still retaining as much intensity in the emission lines as possible, we extracted a spectrum from a position at $\alpha_{\rm J2000}=16^\mathrm{h} $32$^\mathrm{m}$22.58$^\mathrm{s}$; $\delta_{\rm J2000}=-$24\degr28\arcmin32\farcs8, corresponding to an offset of ($-$0.45$\arcsec$ ; $-$0.30$\arcsec$) in the southwestern direction relative to the continuum peak of IRAS16293B. This spectrum, corrected for the LSR velocity (V$_\mathrm{LSR}$ = 2.7 km s$^{-1}$), is used throughout this paper. Figure~\ref{Fig:lines} shows the observed spectra for each of the transitions from Fig.~\ref{Fig:maps}.   

The heavy blending of emission lines at the sensitivity of ALMA complicates the identification and analysis of individual molecular species. For this purpose we therefore calculate synthetic spectra for our target molecules and their physical parameters are derived by fitting synthetic spectra to the data. For the purpose of excluding blended lines from the analysis, we create a reference model containing the synthetic spectrum of emission lines of previously detected complex organic molecules that are expected to be present in the warm gas toward the two sources \citep{Bisschop2008, Jorgensen2011,Jorgensen2012, Jorgensen2016,Coutens2016}. Superimposing the reference model spectrum onto the observed spectrum reveals if a line of interest is blended with any of these species. For our analysis we exclude lines that are severely blended, that is, where the peaks of the emission lines overlap. In addition, we have also checked the lines of interest against other species in the CDMS\footnote{\url{http://www.astro.uni-koeln.de/cdms}} and JPL\footnote{\url{http://spec.jpl.nasa.gov/}} databases \citep{Muller2001,Muller2005,Pickett:1998} with the CASSIS\footnote{\url{http://cassis.irap.omp.eu/}}
software and do not find any clear overlap with any other potential interstellar species.

The synthetic spectra are computed following the approach described in \cite{Goldsmith1999}. We assume that the molecular excitation obeys local thermodynamic equilibrium (LTE), which is reasonable at the densities and scales of the ALMA observations toward IRAS16293B \citep{Jorgensen2016}, and calculate a synthetic spectrum of all transitions from a molecule given a line width, column density, rotational temperature, and source size, assuming Gaussian line profiles. The spectroscopic data for propanal \citep{Butcher1964,Hardy1982,Demaison1987} and ethylene oxide \citep{Cunningham1951,Creswell1974,Hirose1974,Pan1998,Medcraft2012} are available from the CDMS database, while the spectroscopic data for acetone \citep{Groner2002} and acetaldehyde \citep{Kleiner1996} are available from the JPL database.

For the analysis we started by identifying the brightest potential lines of each of the relevant species adopting a full width half maximum (FWHM) line width and the source size remained fixed at 1.0~km s$^{-1}$ and $0.5 \arcsec$, respectively. We then generated a synthetic spectrum by adjusting the temperature and column density ($N_\mathrm{tot}$) until a good fit for those lines was obtained. From this a priori spectrum, we identified approximately ten reasonably non-blended and optically thin $( \tau \le 0.1)$ lines for each species, which we use to minimize the reduced chi-squared statistic:
\begin{equation}
 \chi ^2_\mathrm{red.} =  \frac{1}{N}\sum_{i = 1}^{N} \left(  \frac{ \left(I_{\mathrm{obs}, \, i} - I_{\mathrm{syn}, \, i} \right)}{\sigma_i} \right) ^2 ,
\end{equation}
where $I_\mathrm{obs}$ and $I_\mathrm{syn}$ are the intensities of the observed and synthetic emission lines, respectively, $N$ is the number of lines analyzed and $\sigma$ the RMS error. In the analysis, we varied the column density from $1.0 \times 10^{14}~\mathrm{cm}^{-2}$ -- $1.0 \times 10^{18}~\mathrm{cm}^{-2}$ with small increments and the temperature from 100~K -- 400~K with increments of 25~K, generating a new synthetic spectrum at each increment to evaluate against the observed spectrum at the locations of the chosen lines. Since the emission lines are blended, the reduced $\chi ^2$ is only calculated for the average value of the channels at the very peak of the lines (corresponding to the predicted frequency of the peak $\pm$ 0.25~MHz), instead of over the entire Gaussian bell curve. 

From the reduced $\chi^2$ analysis, acetaldehyde and ethylene oxide show the best fit at $T_\mathrm{ex}$~$\approx$~125~K, while it is difficult to constrain the excitation temperature for propanal and acetone. Our analysis shows that the column densities do not vary greatly with temperature for all species, except for acetone, where a $T_\mathrm{ex}$=400~K results in a column density a factor of ten higher than for $T_\mathrm{ex}$=100~K. A comparison between the synthetic and observed spectrum for acetone reveals that an excitation temperature of approximately 200~K could still be in agreement with the observations, but that a $T_\mathrm{ex}$ of 300~K overproduces some of the lines. Since it appears that the molecules are spatially coexisting and trace the same gas, we therefore assume $T_\mathrm{ex}=125$~K for all molecules. The resulting column densities are summarized in Table~\ref{table:results_combi} and the relative abundance ratios of the different isomers are listed in Table \ref{tcompare}. The uncertainties of $T_\mathrm{ex}$ and $N_\mathrm{tot}$ are dominated by the assumptions that go into the analysis, that is, LTE and Gaussian line profiles, instead of the statistical error. Therefore, the uncertainties are estimated to $\sim 50 \%$ and 25~K on the column density and the emission temperature, respectively.

Figs.~\ref{fig:app_eo_1}--\ref{fig:app_ac_12} in the Appendix show the synthetic spectra of ethylene oxide, propanal, and acetone, respectively, as well as the reference model superimposed on the observed spectrum for all lines where the synthetic spectrum predicts a peak line intensity equal to or above twice the RMS noise of the spectrum. The lines are sorted into descending intensity. We check each line in the synthetic spectra against the observed spectrum for each molecule, and the majority of them provide a reasonable match, within the estimated uncertainty. 
We claim a detection for lines \textit{i}) that are reasonably well separated from other species in the reference model and \textit{ii}) where the integrated line strength over FWHM is larger than three times the statistical uncertainty ($\sqrt{n_\mathrm{chan}} \times \mathrm{RMS}$) of the line and \textit{iii}) where there is a reasonably good fit between the synthetic and the observed spectrum. Table~\ref{tlines} in the appendix lists the spectroscopic catalog values, the integrated intensity over the FWHM for the observed spectrum, and the detection level for the detected lines of ethylene oxide, propanal, and acetone. The transitions are listed with increasing frequency and it should be noted that many of the detected lines are a blend of several internal rotation components.

For ethylene oxide, propanal, and acetone, we detected 20, 44, and 186 lines, respectively. Some of the acetone lines predicted by the models appear to be either slightly shifted or missing. In some cases, this can be explained by the presence of absorption at the same frequency as the predicted lines, but in most cases these lines correspond to transitions with both high K$_{\rm a}$ and low K$_{\rm c}$ quantum numbers (see Table \ref{missing_lines}). None of the missing or shifted lines with high K$_{\rm a}$ and low K$_{\rm c}$ numbers were used for the determination of the spectroscopic parameters. It was admitted by \citet{Groner2002} that these lines do not fit very well. It could be due to perturbations from interactions between the (high K$_{\rm a}$, low K$_{\rm c}$) levels and the levels from the lowest torsional excited states \citep{Groner2002}.

We also search for vinyl alcohol \citep{Saito1976}, another isomer of acetaldehyde and ethylene oxide, but no detection can be claimed so far. With a conservative upper limit of 2\,$\times$\,10$^{15}$ cm$^{-2}$ for the syn form (the lowest energy form of vinyl alcohol), this isomer is less abundant than acetaldehyde and ethylene oxide, similarly to what was found in Sgr B2 by \citet{Belloche2013}.

\begin{table}
\caption{Best fit column densities}             
\label{table:results_combi}      
\centering                        
 \small\addtolength{\tabcolsep}{-1pt} 
\begin{tabular}{l c c }        
\hline\hline                 
Molecule & & $N_\mathrm{tot}$ [cm$^{-2}$]  \bigstrut[t] \\    
\hline
Acetone & \acetone{} & $ 1.7 \times 10^{16}$   \bigstrut[t] \\
Propanal & \propanal{} & $ 2.2 \times 10^{15}$  \bigstrut[t] \\ 
Acetaldehyde & \acetaldehyde{} & $ 7.0 \times 10^{16}$ \bigstrut[t]  \\ 
Ethylene oxide & \ethyleneoxide{} & $ 6.1 \times 10^{15}$ \bigstrut[t] \\
\hline                                   
\end{tabular}
\tablefoot{The results are derived assuming $\theta_\mathrm{source} = 0.5 \arcsec$, $T_\mathrm{ex}=125$~K and FWHM = 1.0 km s$^{-1}$. The column density of propanal was corrected by a factor of 1.489 to take into account the vibrational and conformational contribution at $T$ = 125 K.}
\end{table}

\begin{table*}
 \caption[]{\label{tcompare}Relative abundances in different sources}
 \centering
\begin{tabular}{lccc}
 \hline \hline
Source &  
  \acetone{} / \propanal{} &
   \acetaldehyde{} / \ethyleneoxide{} &
  References  \bigstrut[t] \\
 \hline
IRAS16293-2422 & 8 & 12  & this study \\
Sgr B2(N)  & $\ge 3.6-14.5^{(a)}$  & $3.7-7.4^{(b)}$ & \cite{Belloche2013}\\
Survey of massive SF regions & -- & $1.2-13.2$ & \cite{Ikeda2001}\\
Chemical model: peak gas-phase & 0.22: 0.83: 0.07$^{(c)}$ & -- & \cite{Garrod2013}\\
Chemical model: peak grain-surface & 0.37: 2.3: 0.39$^{(c)}$ & -- & \cite{Garrod2013}\\
Chemical model of hot cores & -- & 1$^{(d)}$ & \cite{Occhiogrosso2014}\\
\hline
\end{tabular}
\tablefoot{$^{(a)}$ Range reflects span for rotational states in the V$_\mathrm{off}$ = 0 km s$^{-1}$ and the V$_\mathrm{off}$ = 10 km s$^{-1}$ components of Sgr B2(N). Propanal is not detected, therefore the upper limit is used after correction for a similar beam filling factor. $^{(b)}$ Range reflects span for the rotational and first torsionally ($\varv_t$=1) excited states of acetaldehyde in the V$_\mathrm{off}$ = -1 km s$^{-1}$ component of Sgr B2(N).$^{(c)}$ Chemical model of hot cores for a {slow},  {medium}, and {fast} model, respectively.$^{(d)}$ The MONACO code (at 200~K and 1.2 $\times$ $10^6$ yrs).}
\end{table*}

\section{Discussion}
\label{sect_discussion}
\begin{figure*}
\centering
\includegraphics[width=16cm]{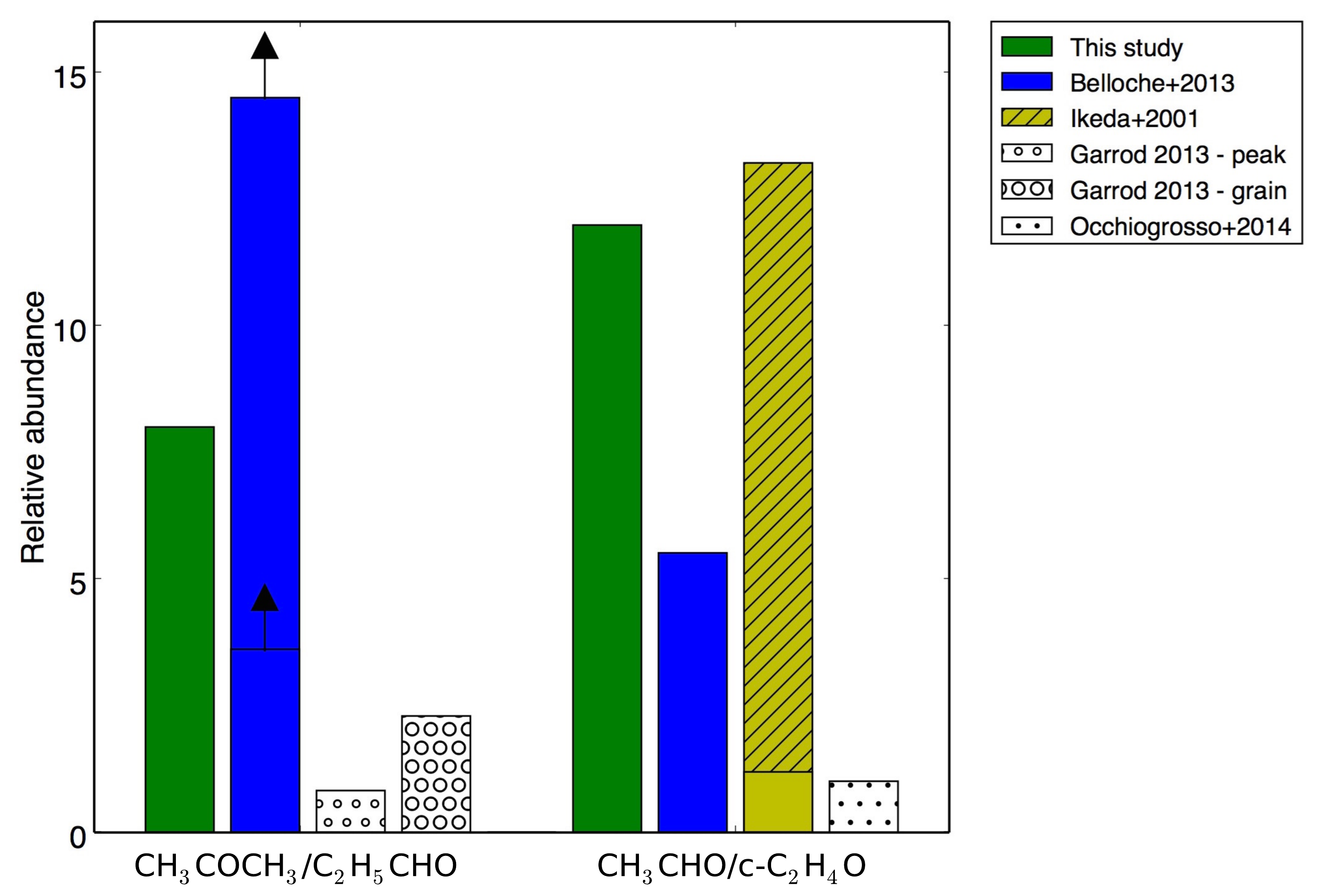}
\caption{Bar plot of the relative abundances of CH$_3$COCH$_3$/ C$_2$H$_5$CHO and CH$_3$CHO/ c-C$_2$H$_4$O from Table \ref{table:results_combi}. The observations are indicated by color bars, while the chemical predictions are shown by white bars with different circle sizes. The two lower limits derived by  \cite{Belloche2013} for CH$_3$COCH$_3$/C$_2$H$_5$CHO are illustrated by upward arrows. The range of CH$_3$CHO/ c-C$_2$H$_4$O ratios determined in ten sources by \cite{Ikeda2001} is indicated by the hatched area. For the CH$_3$CHO/ c-C$_2$H$_4$O ratio from \cite{Belloche2013}, we used the average value of the column densities of the rotational and first torsionally ($v_\mathrm{t}$ = 1) excited states of acetaldehyde. }   
\label{fig:bar}
\end{figure*}

As described in the introduction, the relative abundances of the different isomers are important constraints on chemical models and provide insight into the formation of the complex species. Table \ref{tcompare} lists the different abundance ratios and Fig.~\ref{fig:bar} gives a schematic overview of the entries from the table. A number of different formation pathways have been proposed for the studied species. 

For acetone, the ion-molecule radiative association reaction
\begin{equation}
\rm CH_{3}^{+} + CH_3 CHO \rightarrow (CH_3)_2CHO^{+} + h \nu
,\end{equation} followed by
\begin{equation}
\rm (CH_3)_2CHO^{+} + e^{-} \rightarrow CH_3COCH_3 + H
,\end{equation}
proposed by \citet{Combes1987} has been shown not to be efficient enough to produce the observed values \citep{Herbst1990}. In the model presented by \citet{Garrod2008}, acetone is formed on grains by the addition of CH$_3$ to CH$_3$CO.

\cite{Hollis2004b} proposed the formation of propanal to occur through simple successive hydrogenation:
\begin{equation}
\mathrm{HC}_2\mathrm{CHO} + 2\mathrm{H} \rightarrow \mathrm{CH}_2\mathrm{CHCHO} + 2\mathrm{H} \rightarrow \mathrm{C}_2\mathrm{H}_5\mathrm{CHO}.
\end{equation}
However, \cite{Garrod2013} proposed a different formation route through the addition of HCO and C$_2$H$_5$ radicals on grains. \cite{Garrod2013} found the formation to be most rapid at 30~K, when sublimation of grain-surface methane (CH$_4$) is most efficient. 

Laboratory experiments were conducted by \citet{Bennett2005a,Bennett2005b} to study the synthesis of acetaldehyde, ethylene oxide, and vinyl alcohol in interstellar and cometary ices after irradiation with energetic electrons. Acetaldehyde appeared to be formed in both CO--CH$_4$ and CO$_2$--C$_2$H$_4$ ice mixtures, while ethylene oxide and vinyl alcohol are only detected in CO$_2$--C$_2$H$_4$ ice mixtures \citep{Bennett2005a,Bennett2005b}. While CO, CO$_2$, and CH$_4$ have been observed in interstellar ices, C$_2$H$_4$ is formed as a secondary product by charged particle irradiation and photolysis of CH$_4$ ices and it is therefore likely only present in small concentrations \citep{Bennett2005b} although it may be formed through gas-phase mechanisms under cold, dense conditions. Thus, assuming the relative production rates of acetaldehyde, ethylene oxide and vinyl alcohol are similar, the fractional abundance of acetaldehyde is expected to be higher than that of ethylene oxide and vinyl alcohol \citep{Bennett2005b}.

\subsection{Propanal and acetone}
We have compared our results to predictions from the three-phase (mantle/surface/gas) astrochemical kinetics model, MAGICKAL (Model for Astrophysical Gas and Ice Chemical Kinetics And Layering), as presented in \cite{Garrod2013}. By applying a chemical network to hot-core conditions, the model follows the physico-chemical evolution of a parcel of material from the core from the free-fall collapse of the cloud to the subsequent warm-up phase of the dense core from 8 to 400~K \citep{Garrod2013}. MAGICKAL employs a (modified) rate-equation approach to solve the coupled ice mantle, ice-surface, and gas-phase chemistry allowing radicals on the grains to meet via thermal diffusion at intermediate temperatures and form more complex molecules prior to the complete sublimation of the dust-grain ice at higher temperatures. \cite{Garrod2013} uses three different warm-up models: \textit{fast}, \textit{medium} and \textit{slow}. Here we compare our results to all three models, but note that the fast warm-up model should, in principle, be the best match to the observations because the time for this model to reach 200~K is 5 $\times \, 10^4$ yr which is comparable to the dynamical age of $\sim$ 1--3 $\times \, 10^{4}$~yr for IRAS16293 as derived by \cite{Schoier2002}. 

\cite{Garrod2013} finds relative peak gas-phase abundances of \acetone{}/~\propanal{} of 0.22, 0.83, and 0.07 for the fast, medium and slow model, respectively. All three models predict a higher abundance of propanal compared to acetone, which is the opposite trend of our ratio of eight. Also, the upper limit toward Sgr~B2(N) reported by \cite{Belloche2013} translates into a lower limit for \acetone{}/\propanal{} of 3.6, which is consistent with our findings. One explanation may be that the model of \cite{Garrod2013} uses a relatively low binding energy for acetone (3500~K), producing a desorption temperature of approximately 70~K. As discussed by \cite{Garrod2008}, this low-temperature desorption results in rapid destruction of acetone in the gas-phase. Our observational fit to the excitation temperature of 125~K suggests that acetone is more likely desorbed from grains at the higher temperatures more commonly associated with complex organics, which would allow the majority of grain-surface formed acetone to survive for a significant period in the gas phase. 

If we compare the peak {\em grain-surface} abundances of acetone and propanal produced in the \cite{Garrod2013} chemical model which would be more representative of this situation, ratios of 0.37, 2.3, and 0.39  are obtained, respectively.
The quantities of acetone and propanal produced on grains in the model are, in the case of the intermediate warm-up timescale, only a factor of a few below the observed ratio. However, it should be borne in mind that the efficient production of acetone depends, in this model, on the rate at which the CH$_3$CO radical may be produced on the grains. This may be achieved either through direct photodissociation of CH$_3$CHO or by the abstraction of a H-atom from this molecule by OH or NH$_2$. The rates of each of these processes are not well defined by experiment, and these uncertainties could easily induce a variation in acetone production of a few factors. It is also likely that the physical conditions, which in the \cite{Garrod2013} model are generic, representative hot-core conditions, may not be accurate for the specific case of IRAS16293.

\subsection{Ethylene oxide and acetaldehyde}
\cite{Ikeda2001} searched for acetaldehyde and ethylene oxide in several massive star-forming regions. They detect both molecules in ten sources and find \acetaldehyde{}/\ethyleneoxide{} spanning a range from 1.2 in Sgr B2(N) to 13.2 in W51e1/e2. \cite{Belloche2013} also observed these molecules towards Sgr B2(N) and found a slightly higher value than \cite{Ikeda2001} of 3.7--7.4. It thus seems that our observed value of 12 in a low-mass YSO is toward the high end of the range observed in these high-mass regions, but that source-to-source variations may be larger than between the different groups of sources.

\cite{Occhiogrosso2014} used a two-stage (grain/gas) model, MONACO, to predict the gaseous acetaldehyde and ethylene oxide abundances during the cooling-down and subsequent warm-up phase of a hot core. At 200~K and $1.2 \times 10^{6}$ yrs, the fractional abundance of ethylene oxide and acetaldehyde with respect to total H is $2 \times 10^{-9}$ for both molecules, which means that the relative abundance between the two species is unity. As previously mentioned, based on their laboratory experiments, \cite{Bennett2005a,Bennett2005b} expect the relative abundance of \acetaldehyde{} / \ethyleneoxide{} to be larger than unity. Again, it seems that there are some variations in the observed acetaldehyde--to--ethylene oxide ratios, and that the model results of \cite{Occhiogrosso2014} best reproduce the lower end in that range, while our measurements are at the opposite end, more than an order of magnitude above. Nevertheless, given the variations seen in the models for acetone and propanal,  whether the specific physical structures of the sources can be part of the explanation remains to be explored.

\section{Conclusion}
\label{sect_conclu}
We have carried out the first investigation of the oxygen bearing species in the ALMA PILS survey of the protostellar binary system IRAS16293. Our main findings are summarized as follows:

\begin{enumerate}
\item We have detected the molecules ethylene oxide (\ethyleneoxide{}), acetone (\acetone{}), and propanal (\propanal{}) for the first time toward a solar-type protostar. We have verified that the emission of these species, along with acetaldehyde (\acetaldehyde{}), originates from the compact central region of the protostar, which confirms our assumption that these molecules spatially coexist. We determined a common excitation temperature, $T_{\rm ex} \approx 125$~K for all four molecules and use this to determine column densities for each species. 

\item Compared to previous observations, our results for the relative abundance ratio of \acetone{}/\propanal{} are consistent with the lower limit found by \cite{Belloche2013} of SgrB2(N). The ratio for \acetaldehyde{}/\ethyleneoxide{} is comparable to the largest value in the span of observed values of high-mass sources from \cite{Ikeda2001} (variation between the sources in that sample of approximately an order of magnitude). This suggests that the chemistry in the most central part of IRAS16293 (the hot corino region) is not significantly different from those of the high-mass hot cores, but that there may still be measurable source-to-source variations.

\item Contrary to our result, the models in \cite{Garrod2013} predict propanal to be more abundant than acetone, except for the peak grain-surface abundances in the medium warm-up model, where the prediction is only few factors different from our result. \cite{Occhiogrosso2014} find the ratio of \acetaldehyde{}/\ethyleneoxide{} to be unity which is consistent with the lowest observed value of a high-mass star forming region (\cite{Ikeda2001}). All of the models investigated here return low relative abundances compared to our results, but they are however in reasonable agreement with the lowest value in the ranges reported by \cite{Ikeda2001} and \cite{Belloche2013}.

\end{enumerate}

The results from this paper imply that although the chemical models can reproduce the observations for some high-mass protostars reasonably
well, they need to be modified to reflect the observed range of values for high-mass sources as well as our low-mass source. As discussed, the models would improve with better-defined reaction rates while including more species in the chemical networks could also improve model predictions. More observations, in particular toward low-mass sources, are needed for comparison with models to further constrain the formation pathways.   

The detections also demonstrate the great potential of spectral surveys such as PILS for identifying new species that have so far gone undetected toward solar-type stars. New detections of complex organic molecules and the determination of their relative abundances for the first time in a solar-type protostar is important because it substantiates the chemical complexity of IRAS16293 and can be used to constrain astrochemical models. The relative abundances reveal information of the formation pathway of the molecules and enable comparisons with models and laboratory experiments. In addition, the comparison of the ratios found in high-mass sources and low-mass protostars is vital to understanding the environmental effects on the formation of different molecular species.

\begin{acknowledgements}
This research was made possible through a Lundbeck Foundation Group Leader Fellowship as well as the European Research Council (ERC) under the European Union Horizon 2020 research and innovation programme (grant agreement No 646908) through ERC Consolidator Grant ``S4F'' to JKJ. Research at Centre for Star and Planet Formation is funded by the Danish National Research Foundation. The work of A.C. was funded by the STFC grant ST/M001334/1. A.C. thanks the COST action CM1401 Our Astrochemical History for additional financial support. RTG acknowledges the support of the NASA APRA program, though grant NNX15AG07G. Astrochemistry in Leiden is supported by the European Union A-ERC grant 291141 CHEMPLAN, by the Netherlands Research School for Astronomy (NOVA), by a Royal Netherlands Academy of Arts and Sciences (KNAW) professor prize. The research leading to these results has received funding from the European Commission Seventh Framework Programme (FP/2007-2013) under grant agreement No 283393 (RadioNet3).

This paper makes use of the following ALMA data: ADS/JAO.ALMA\#2013.1.00278.S. ALMA is a partnership of ESO (representing its member states), NSF (USA) and NINS (Japan), together with NRC (Canada) and NSC and ASIAA (Taiwan), in cooperation with the Republic of Chile. The Joint ALMA Observatory is operated by ESO, AUI/NRAO and NAOJ. 

\end{acknowledgements}


\bibliographystyle{aa}
\bibliography{new}
\clearpage
\begin{appendix}

\section{Observed and synthetic spectra}

\begin{figure*}
\centering
\includegraphics[width=0.9\textwidth]{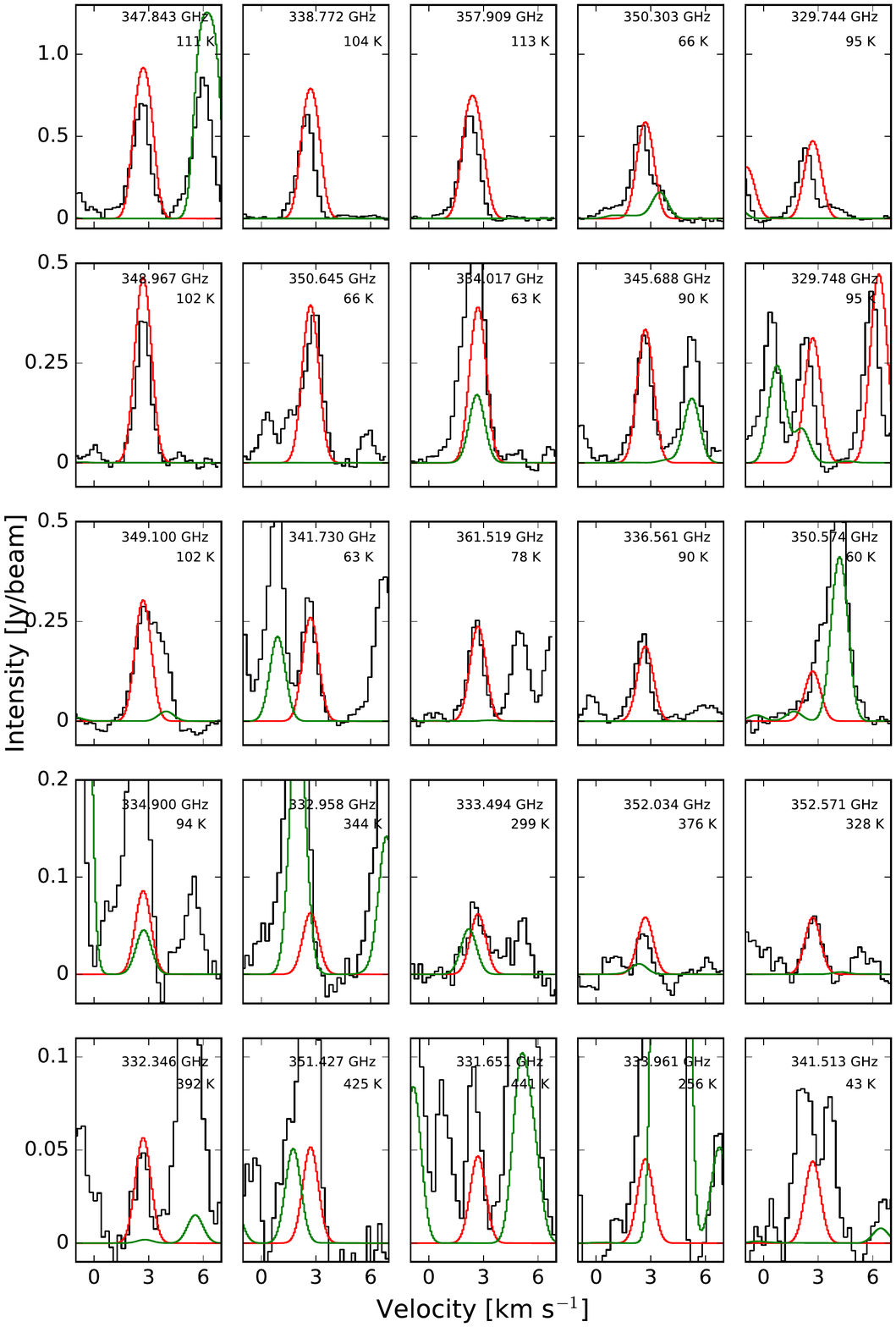}
\caption{Ethylene oxide (\ethyleneoxide{}): Synthetic spectrum in red and reference model in green superimposed onto observed spectrum.}
\label{fig:app_eo_1}
\end{figure*}
\begin{figure*}
\centering
\includegraphics[width=0.9\textwidth]{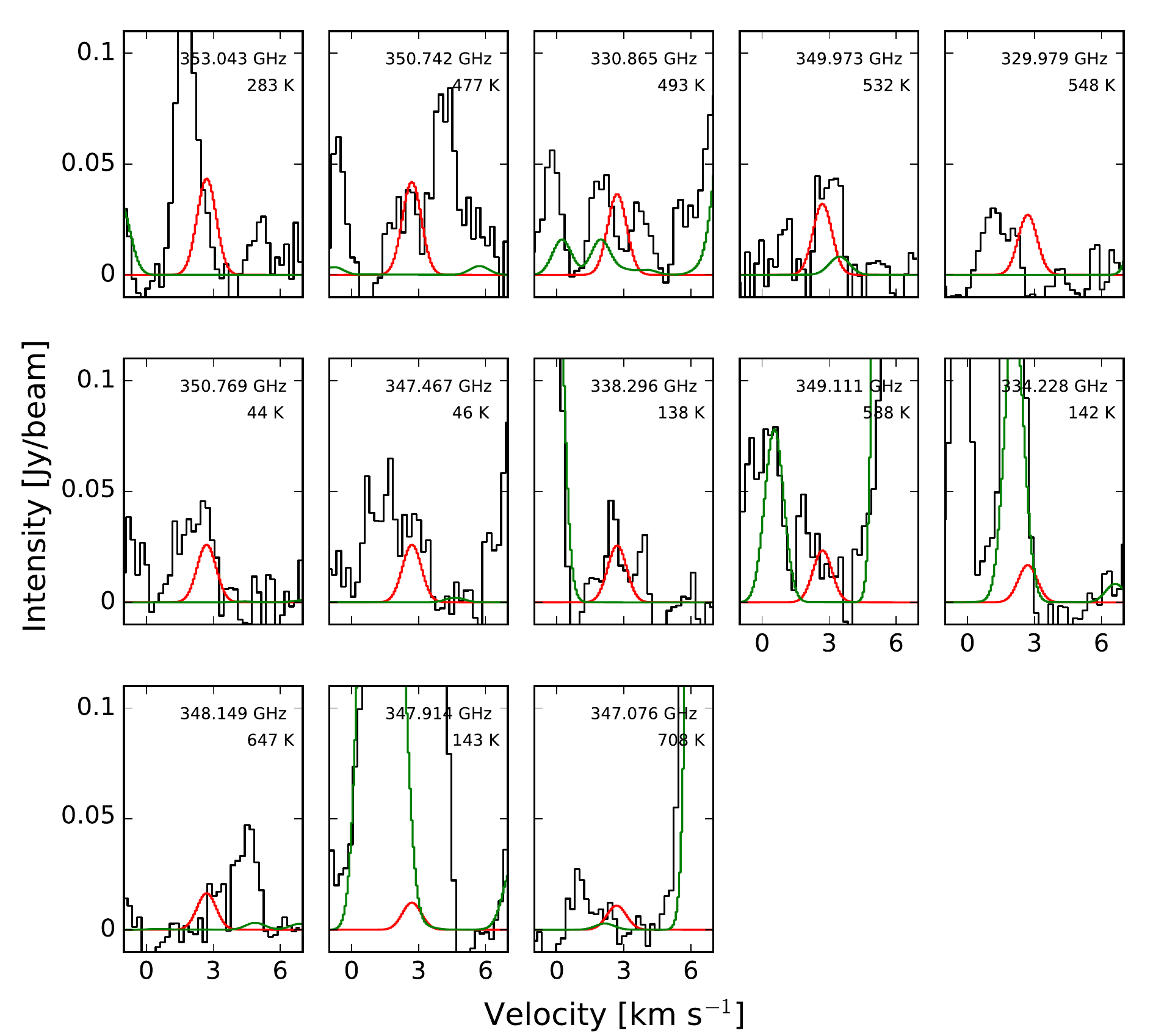}
\caption{Ethylene oxide (\ethyleneoxide{}): Synthetic spectrum in red and reference model in green superimposed onto observed spectrum.}
\label{fig:app_eo_2}
\end{figure*}

\begin{figure*}
\centering
\includegraphics[width=0.9\textwidth]{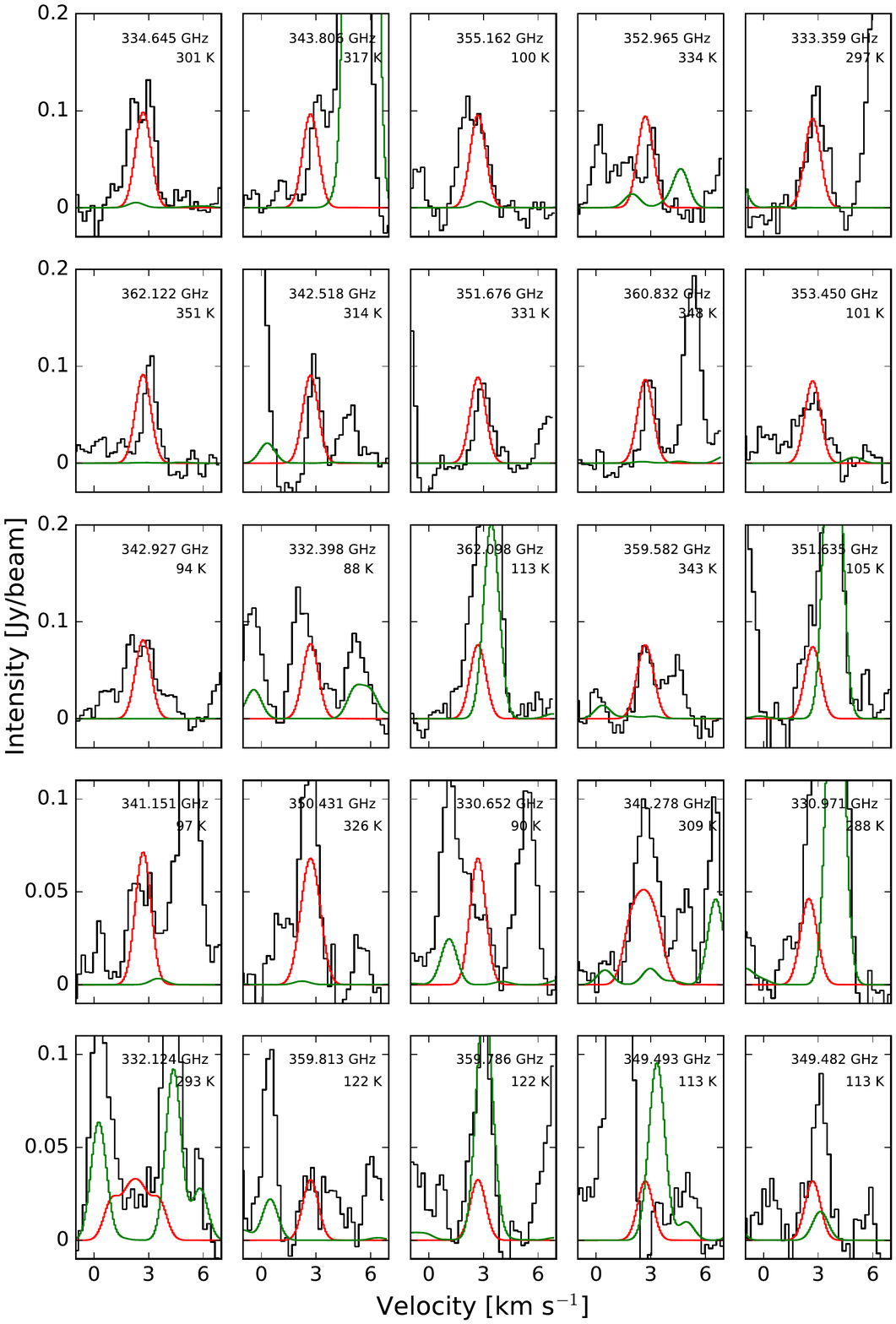}
\caption{Propanal (\propanal{}): Synthetic spectrum in red and reference model in green superimposed onto observed spectrum.}
\label{fig:app_pr_1}
\end{figure*}
\begin{figure*}
\centering
\includegraphics[width=0.9\textwidth]{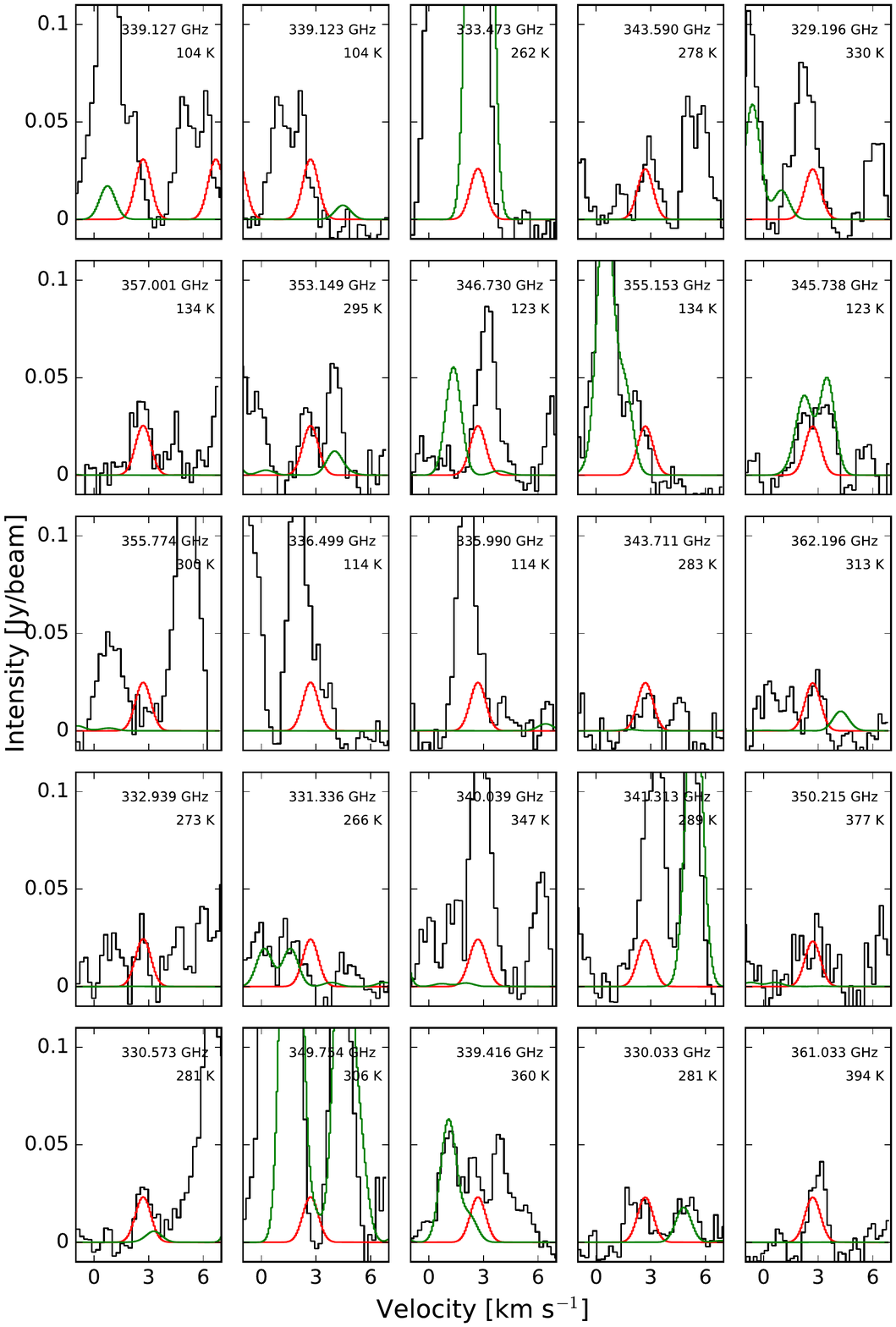}
\caption{Propanal (\propanal{}): Synthetic spectrum in red and reference model in green superimposed onto observed spectrum.}
\label{fig:app_pr_2}
\end{figure*}
\begin{figure*}
\centering
\includegraphics[width=0.9\textwidth]{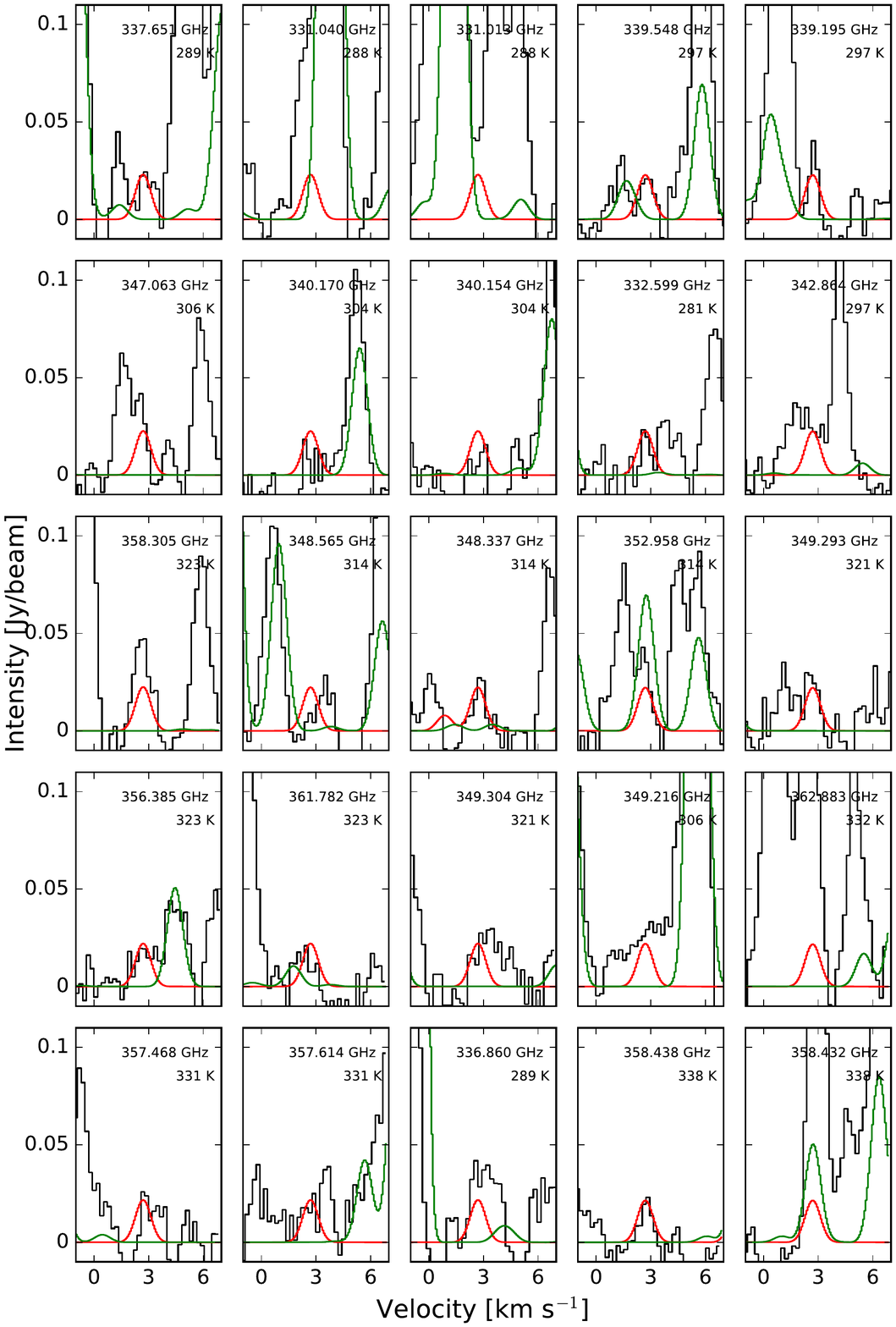}
\caption{Propanal (\propanal{}): Synthetic spectrum in red and reference model in green superimposed onto observed spectrum.}
\label{fig:app_pr_3}
\end{figure*}
\begin{figure*}
\centering
\includegraphics[width=0.9\textwidth]{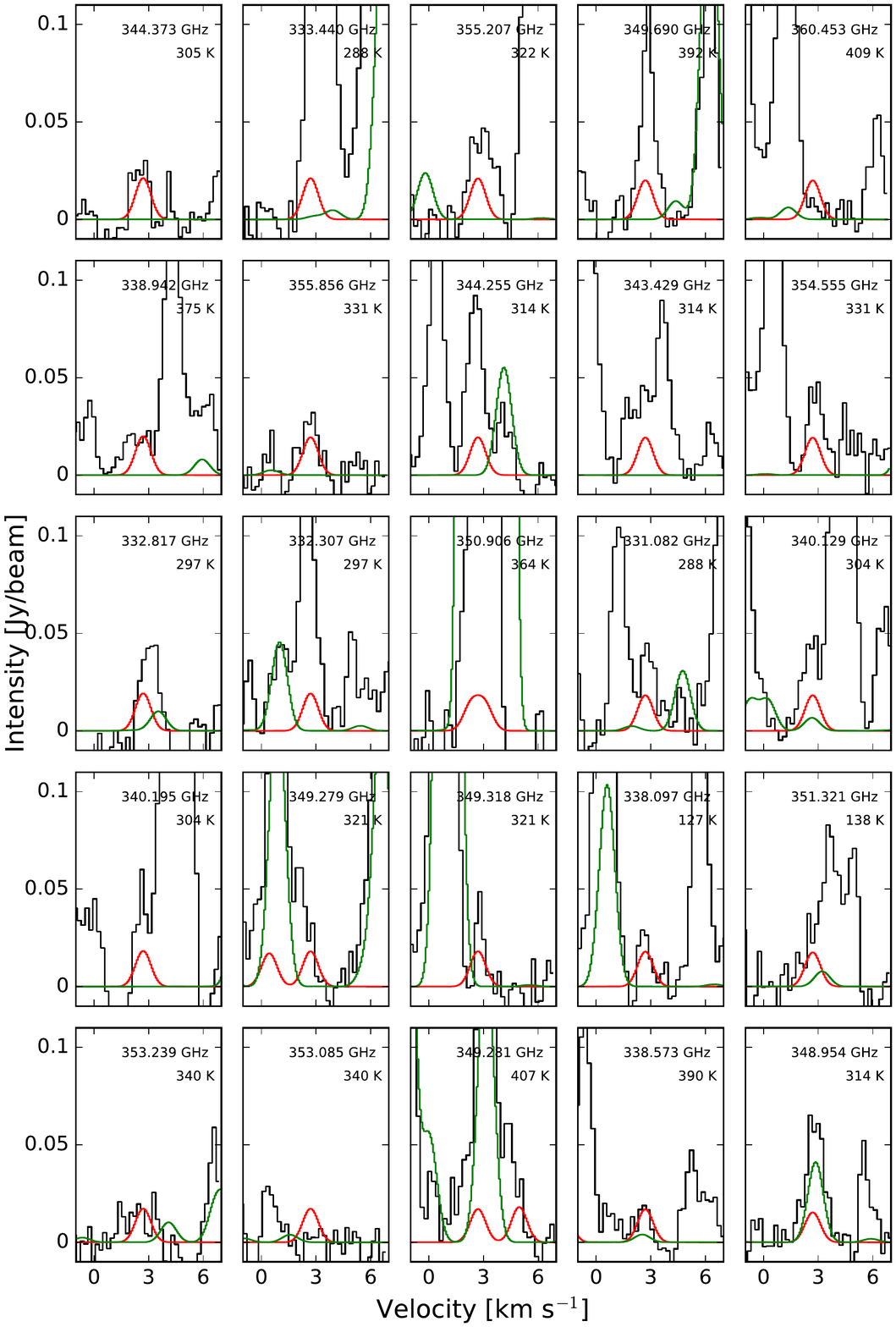}
\caption{Propanal (\propanal{}): Synthetic spectrum in red and reference model in green superimposed onto observed spectrum.}
\label{fig:app_pr_4}
\end{figure*}
\begin{figure*}
\centering
\includegraphics[width=0.9\textwidth]{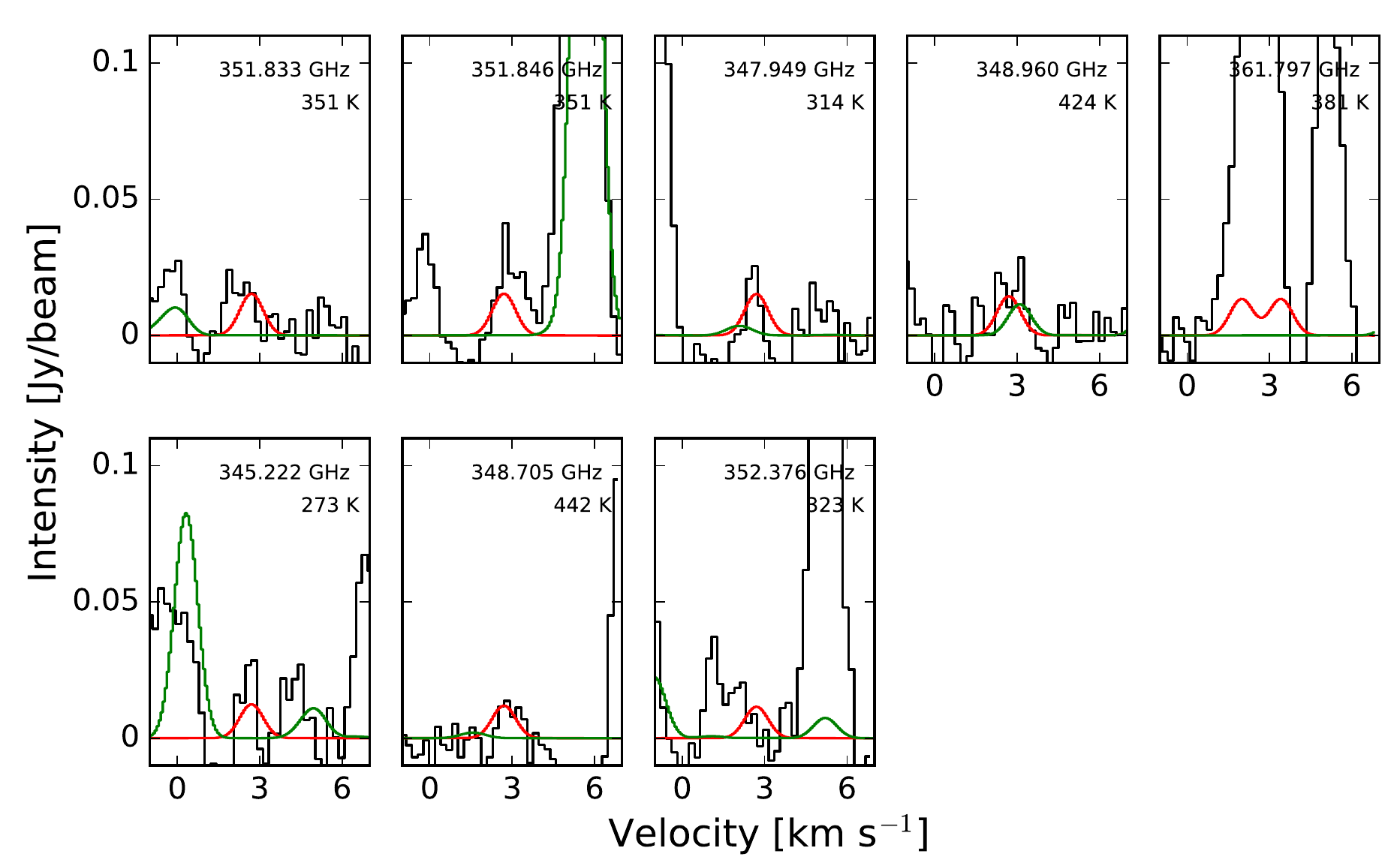}
\caption{Propanal (\propanal{}): Synthetic spectrum in red and reference model in green superimposed onto observed spectrum.}
\label{fig:app_pr_5}
\end{figure*}

\begin{figure*}
\centering
\includegraphics[width=0.9\textwidth]{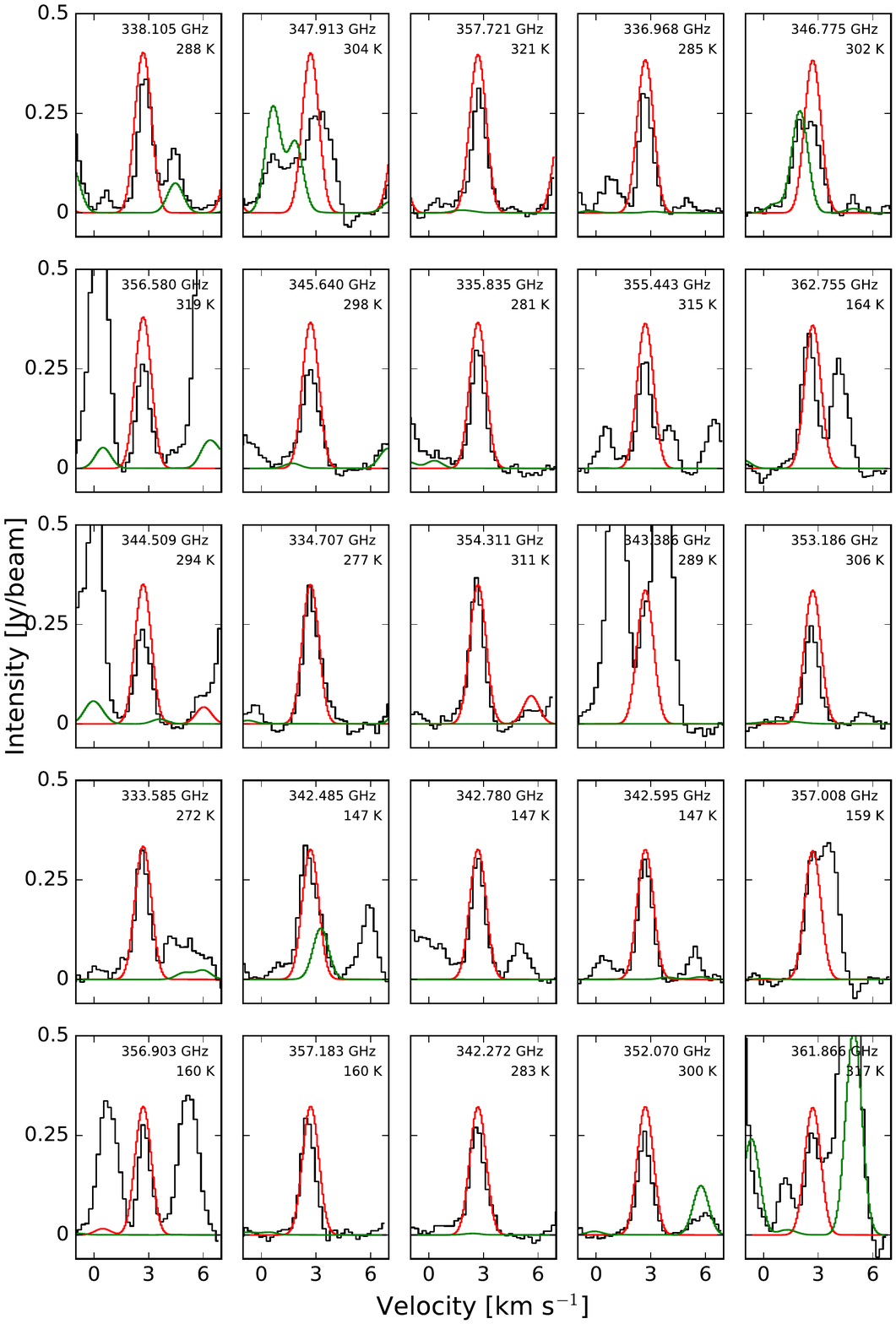}
\caption{Acetone (\acetone{}): Synthetic spectrum in red and reference model in green superimposed onto observed spectrum.}
\label{fig:app_ac_1}
\end{figure*}

\begin{figure*}
\centering
\includegraphics[width=0.9\textwidth]{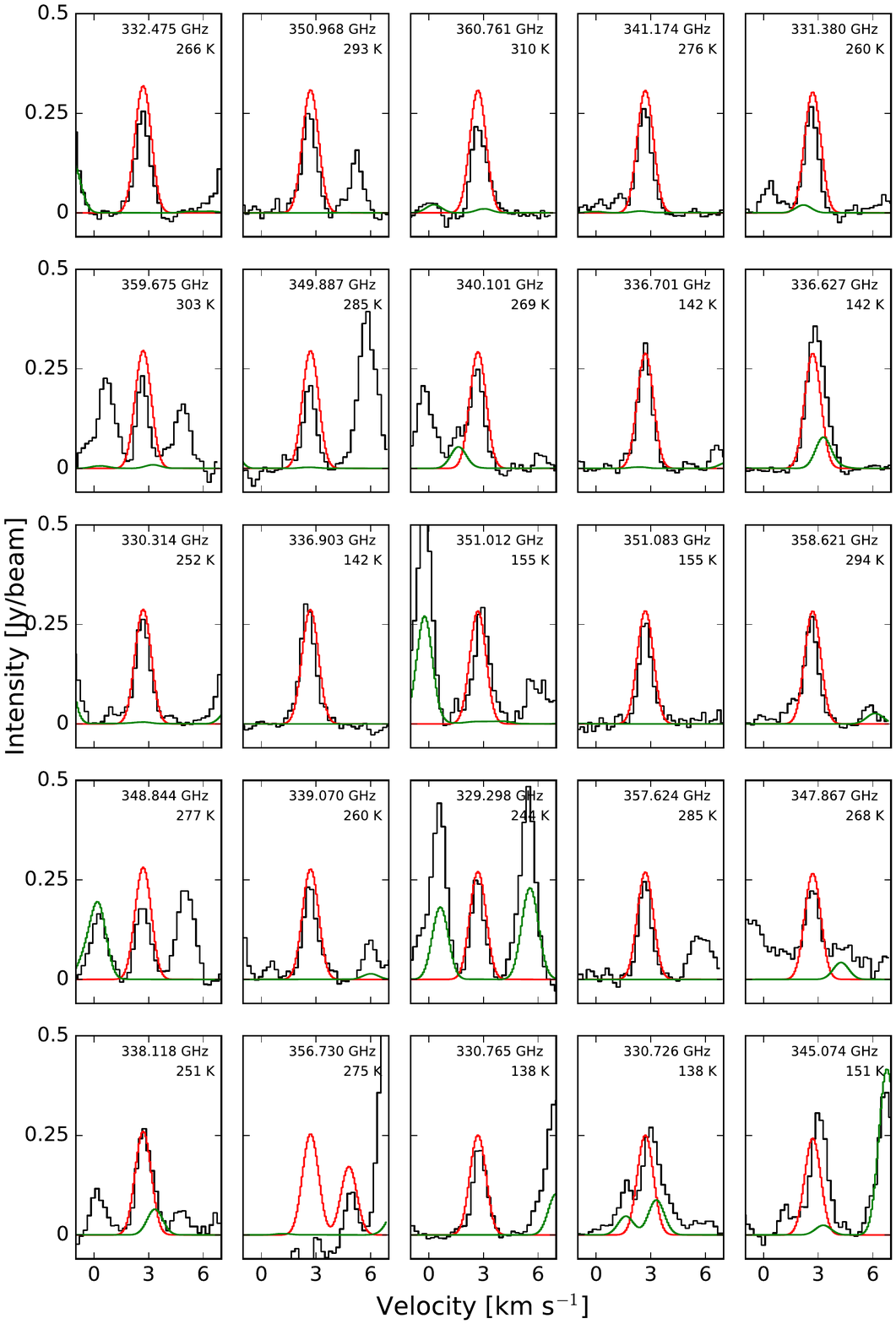}
\caption{Acetone (\acetone{}): Synthetic spectrum in red and reference model in green superimposed onto observed spectrum.}
\label{fig:app_ac_2}
\end{figure*}

\begin{figure*}
\centering
\includegraphics[width=0.9\textwidth]{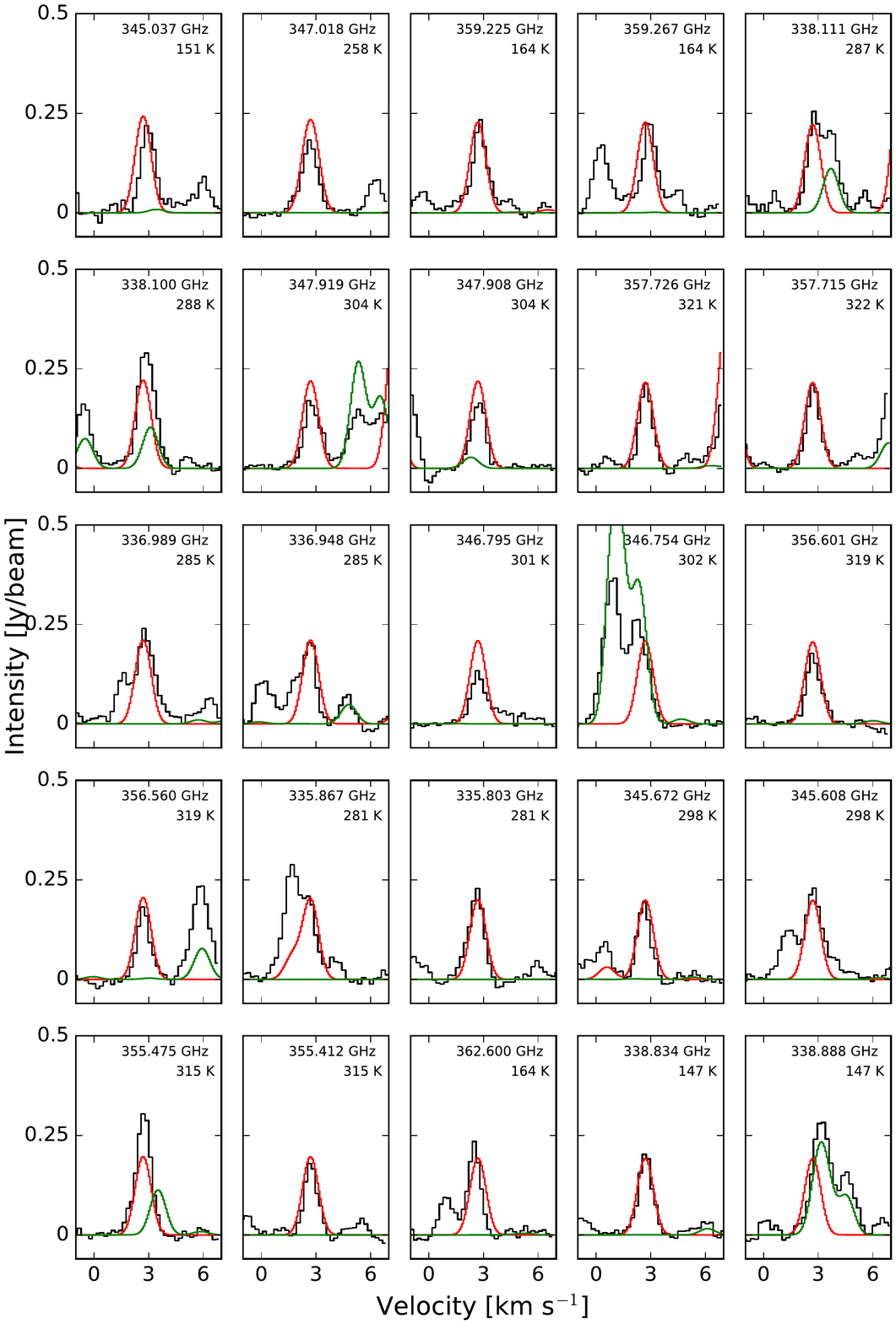}
\caption{Acetone (\acetone{}): Synthetic spectrum in red and reference model in green superimposed onto observed spectrum.}
\label{fig:app_ac_3}
\end{figure*}

\begin{figure*}
\centering
\includegraphics[width=0.9\textwidth]{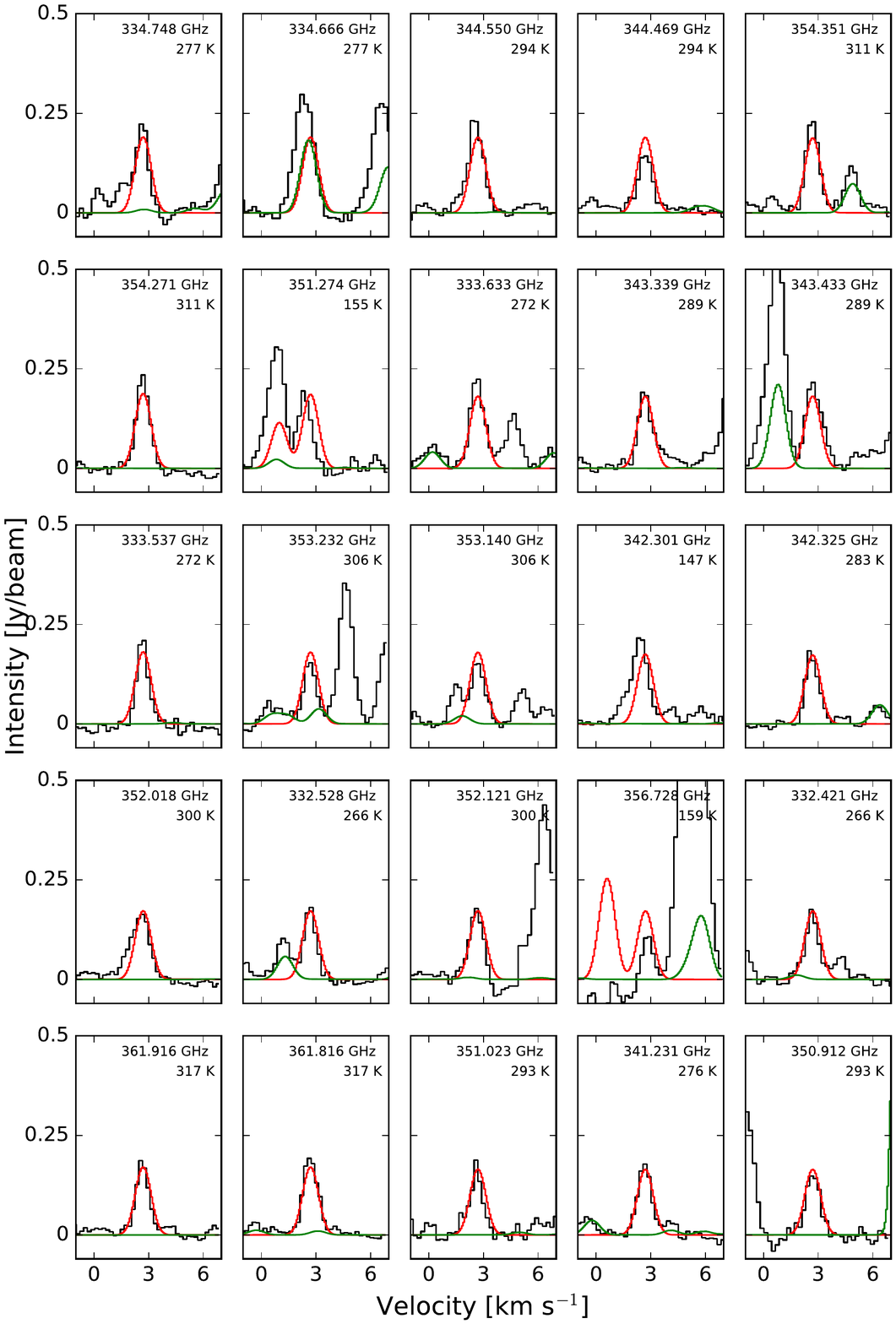}
\caption{Acetone (\acetone{}): Synthetic spectrum in red and reference model in green superimposed onto observed spectrum.}
\label{fig:app_ac_4}
\end{figure*}

\begin{figure*}
\centering
\includegraphics[width=0.9\textwidth]{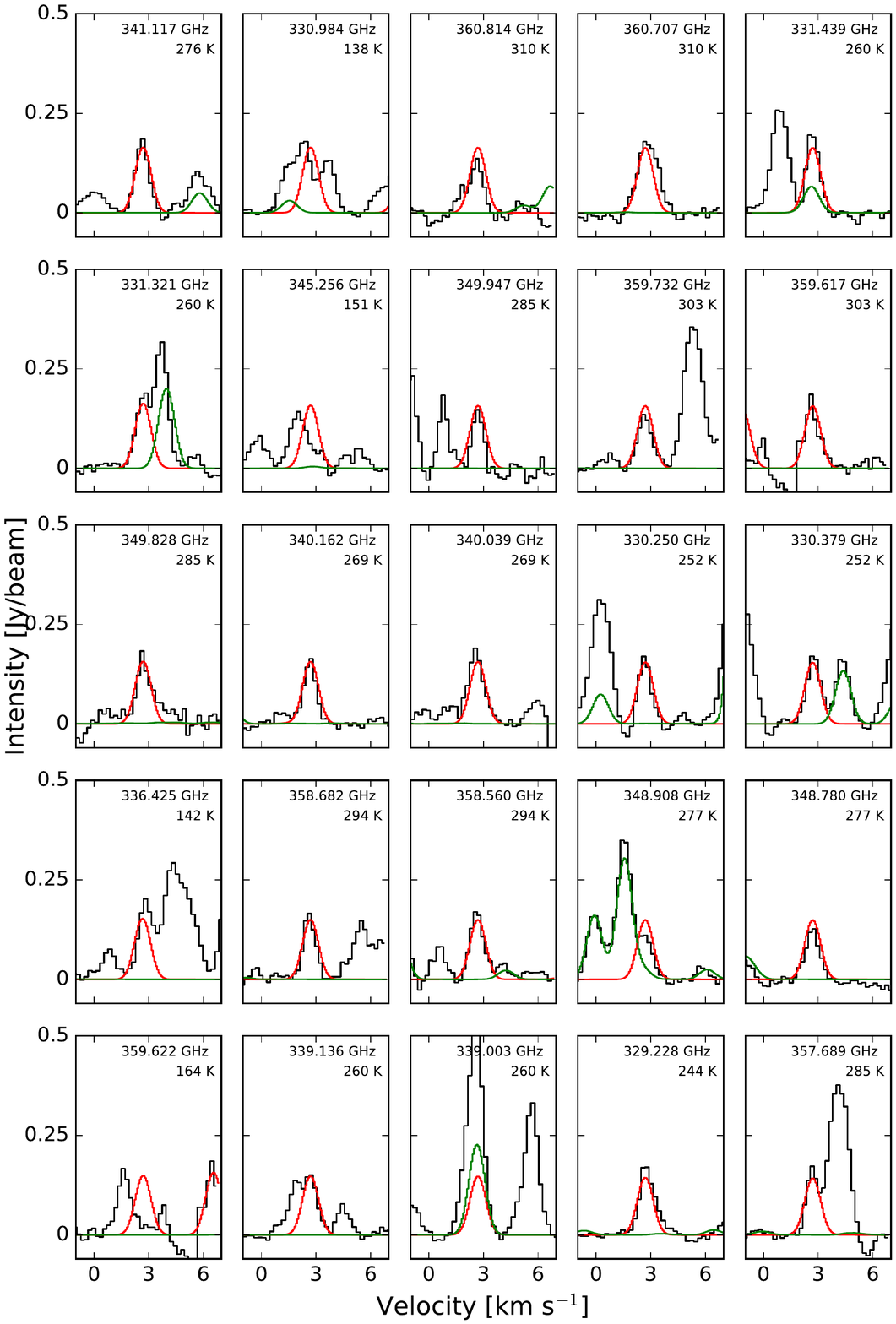}
\caption{Acetone (\acetone{}): Synthetic spectrum in red and reference model in green superimposed onto observed spectrum.}
\label{fig:app_ac_5}
\end{figure*}

\begin{figure*}
\centering
\includegraphics[width=0.9\textwidth]{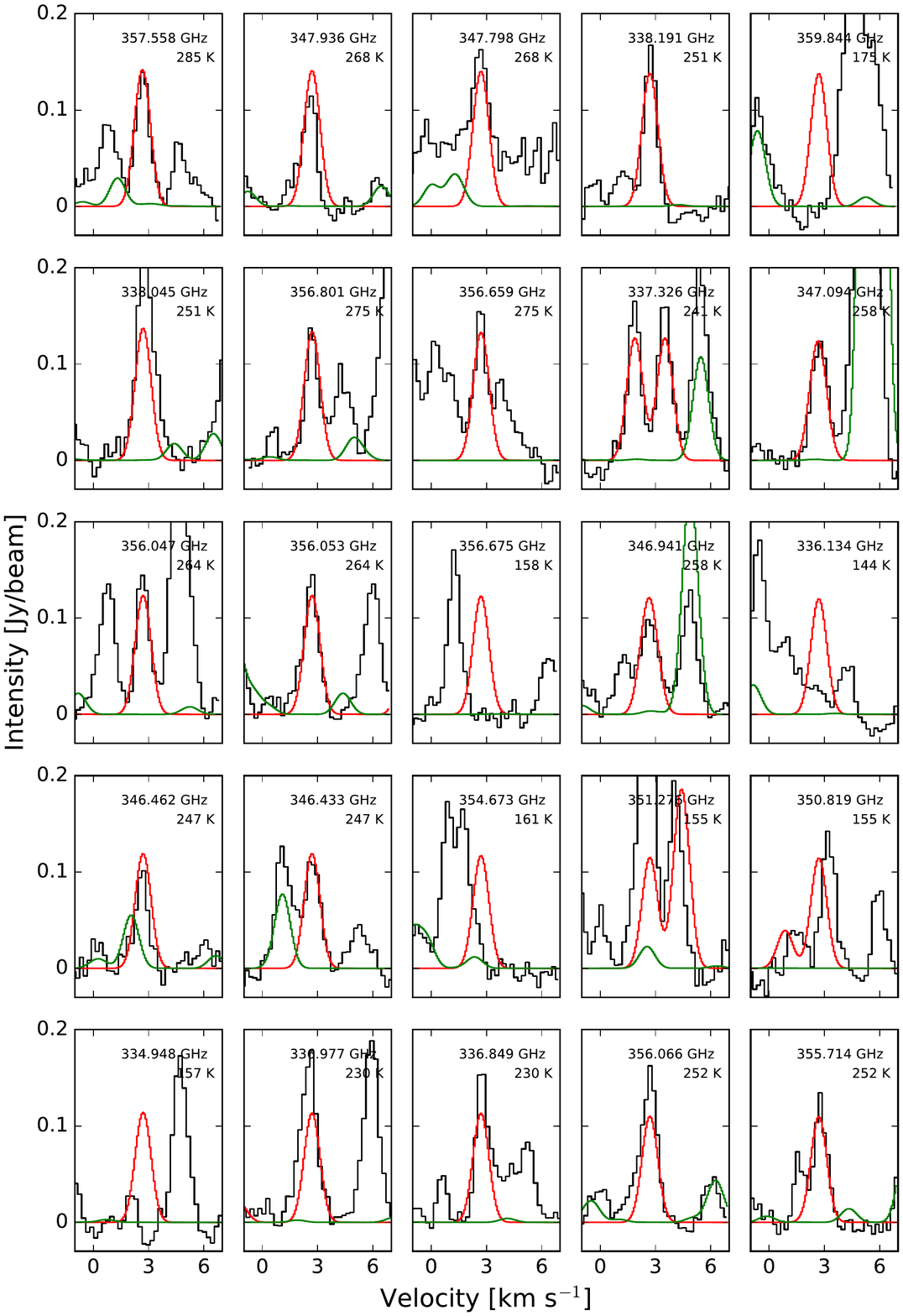}
\caption{Acetone (\acetone{}): Synthetic spectrum in red and reference model in green superimposed onto observed spectrum.}
\label{fig:app_ac_6}
\end{figure*}

\begin{figure*}
\centering
\includegraphics[width=0.9\textwidth]{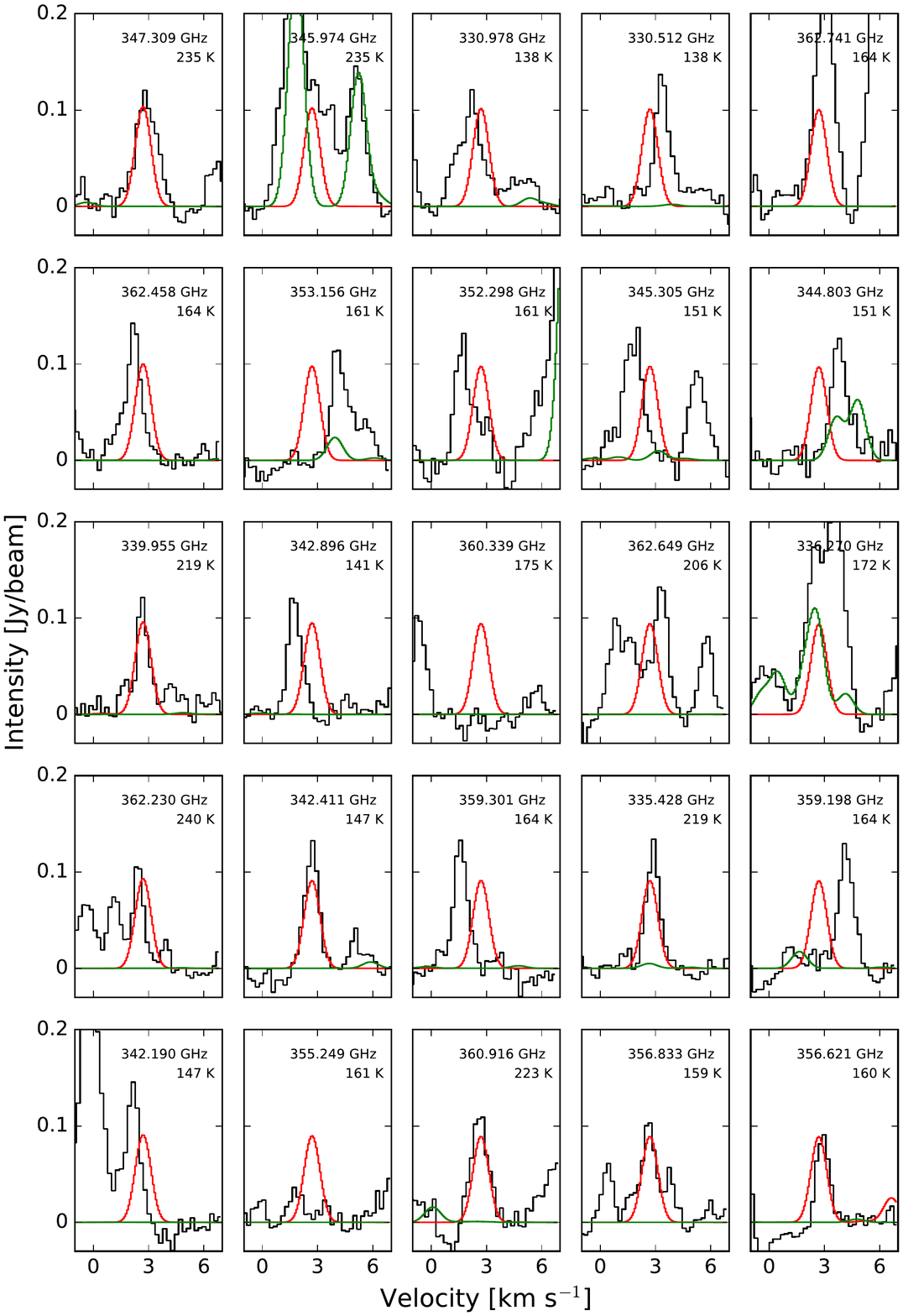}
\caption{Acetone (\acetone{}): Synthetic spectrum in red and reference model in green superimposed onto observed spectrum.}
\label{fig:app_ac_7}
\end{figure*}

\begin{figure*}
\centering
\includegraphics[width=0.9\textwidth]{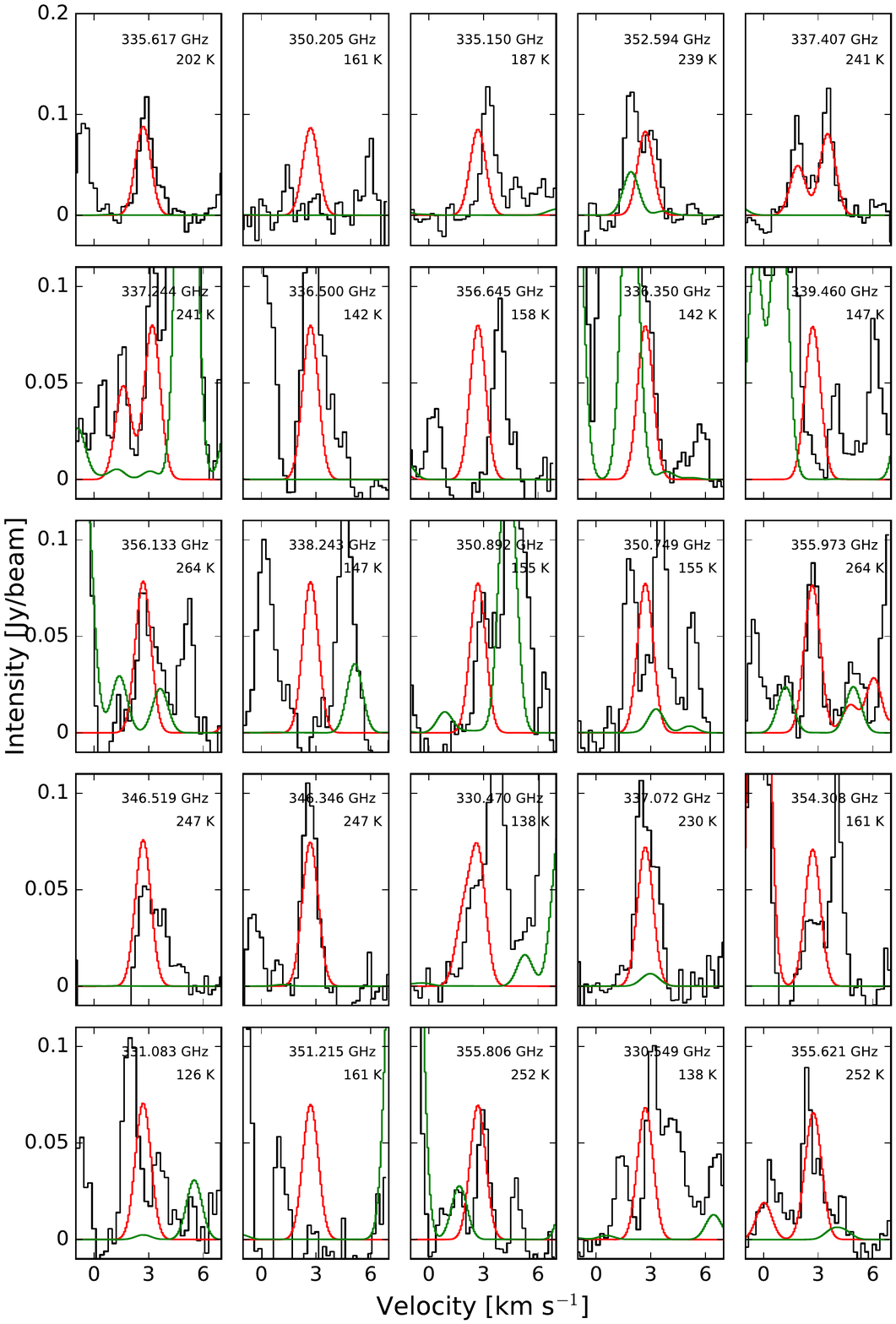}
\caption{Acetone (\acetone{}): Synthetic spectrum in red and reference model in green superimposed onto observed spectrum.}
\label{fig:app_ac_8}
\end{figure*}

\begin{figure*}
\centering
\includegraphics[width=0.9\textwidth]{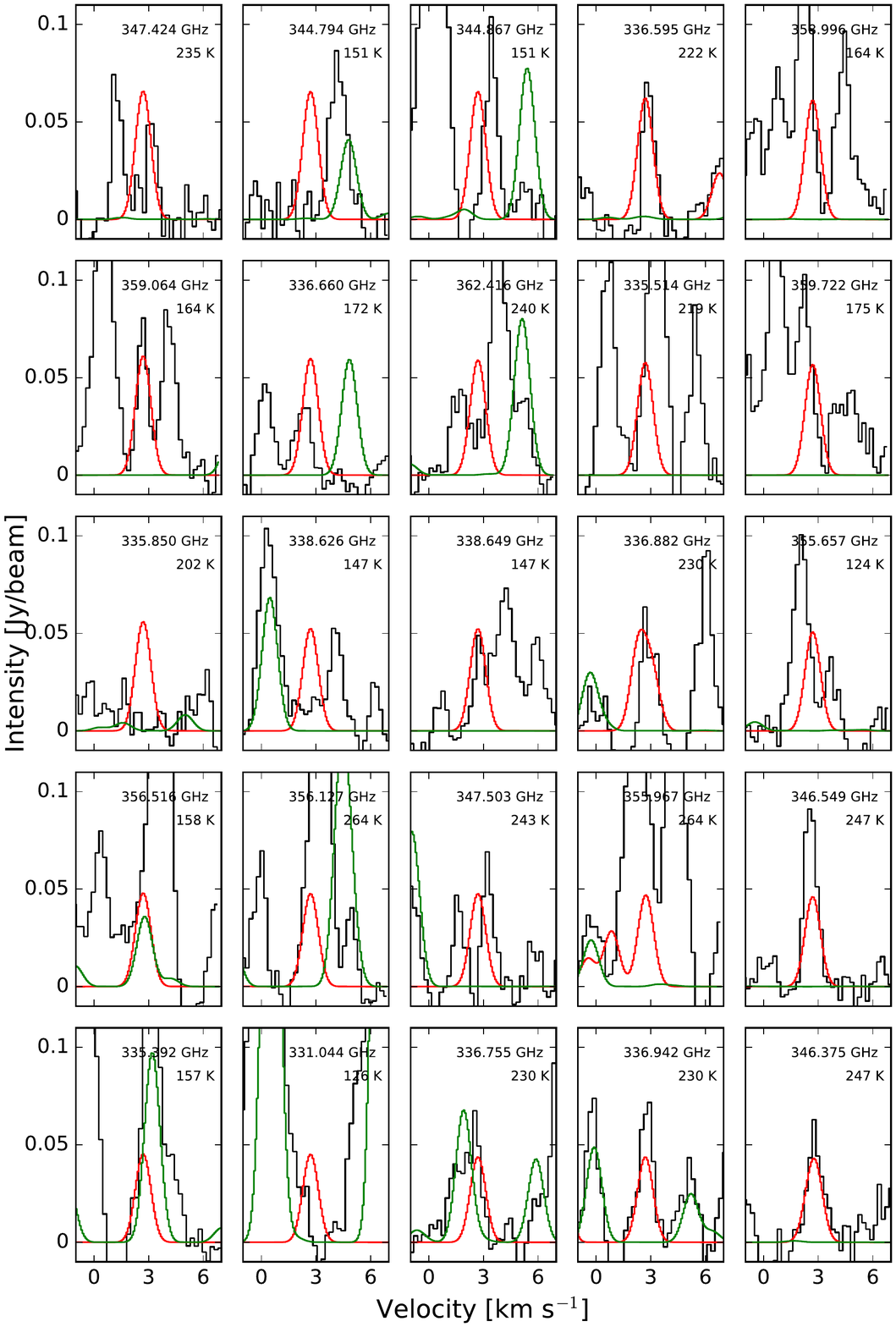}
\caption{Acetone (\acetone{}): Synthetic spectrum in red and reference model in green superimposed onto observed spectrum.}
\label{fig:app_ac_9}
\end{figure*}

\begin{figure*}
\centering
\includegraphics[width=0.9\textwidth]{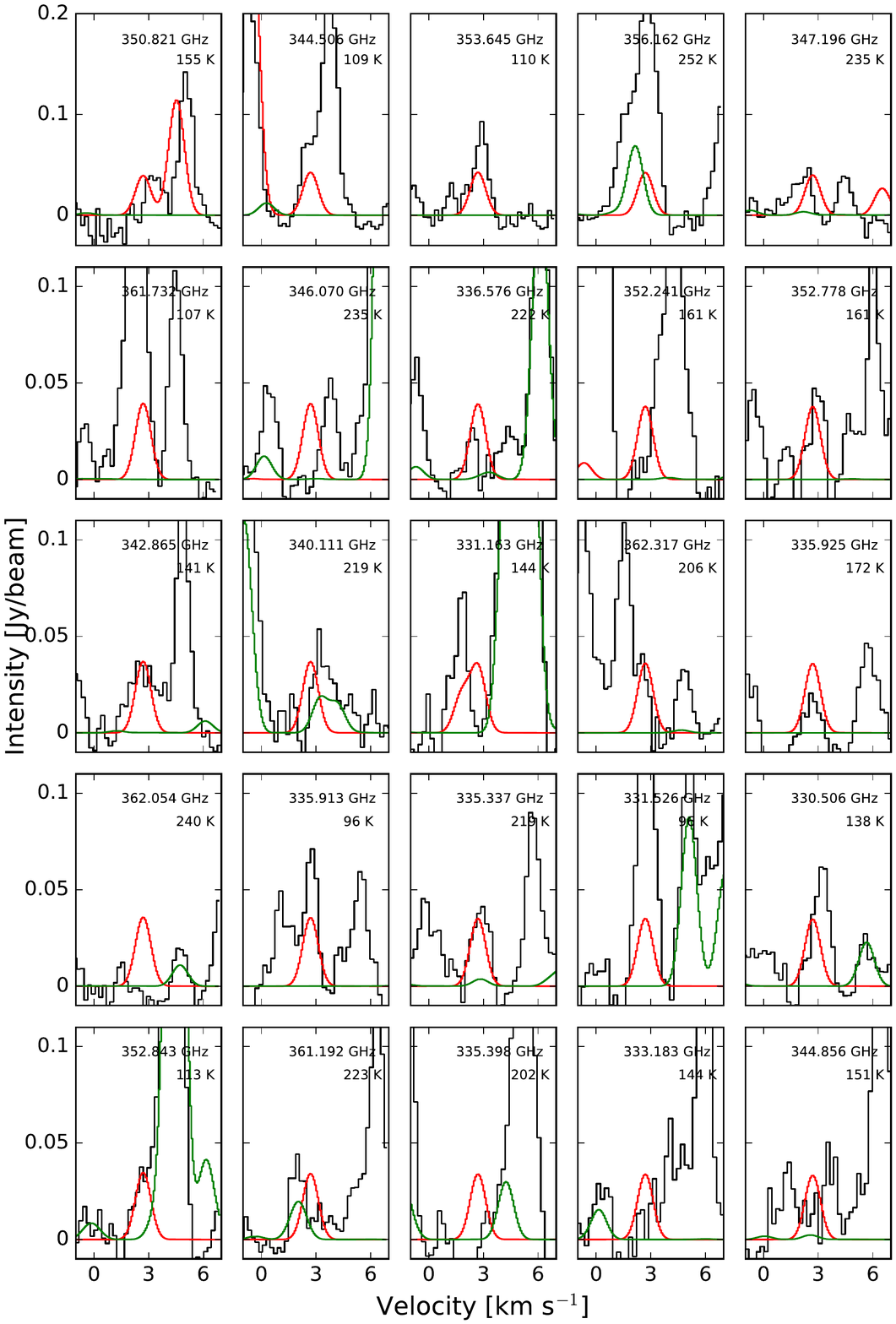}
\caption{Acetone (\acetone{}): Synthetic spectrum in red and reference model in green superimposed onto observed spectrum.}
\label{fig:app_ac_10}
\end{figure*}

\begin{figure*}
\centering
\includegraphics[width=0.9\textwidth]{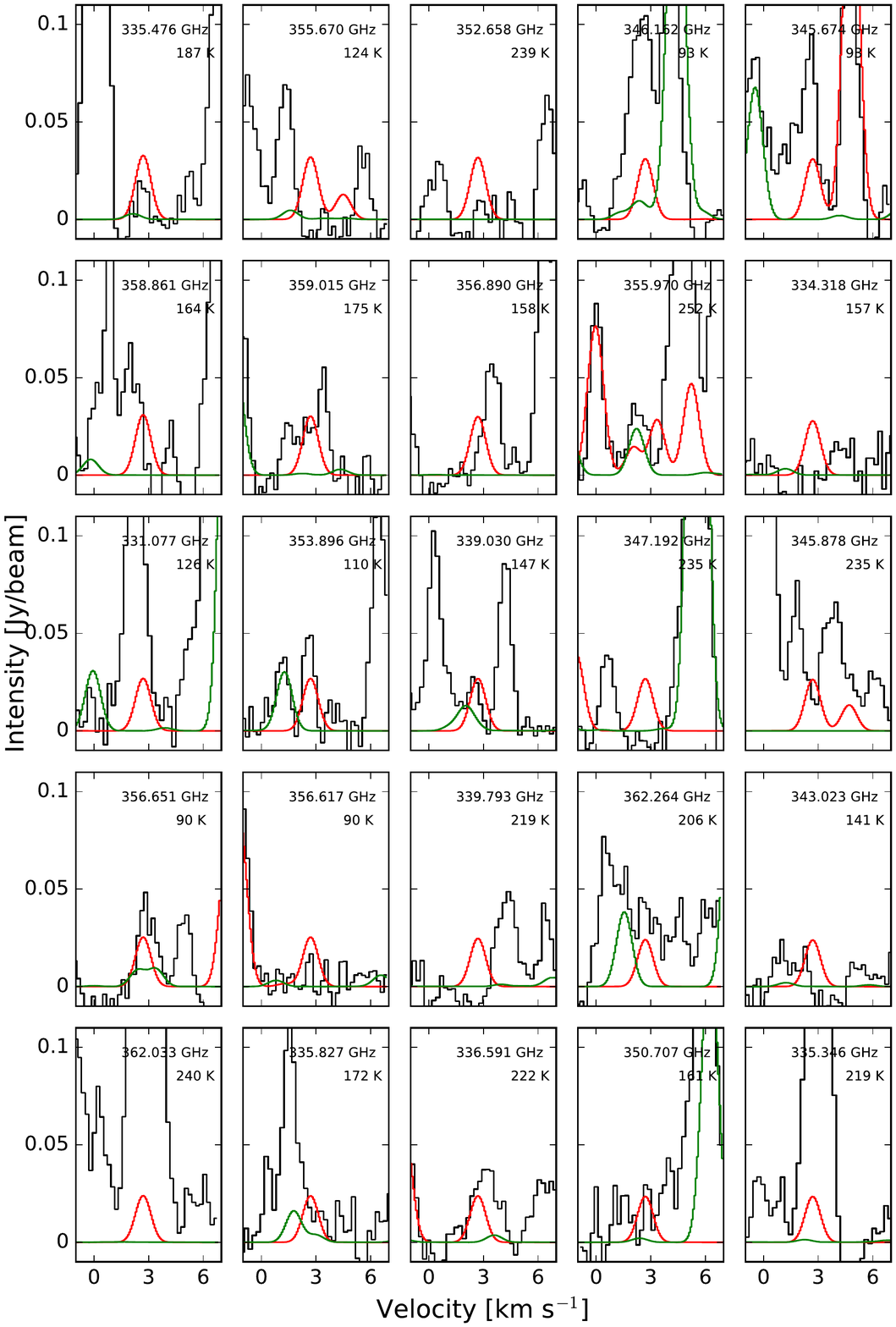}
\caption{Acetone (\acetone{}): Synthetic spectrum in red and reference model in green superimposed onto observed spectrum.}
\label{fig:app_ac_11}
\end{figure*}

\clearpage
\begin{figure*}
\centering
\includegraphics[width=0.9\textwidth]{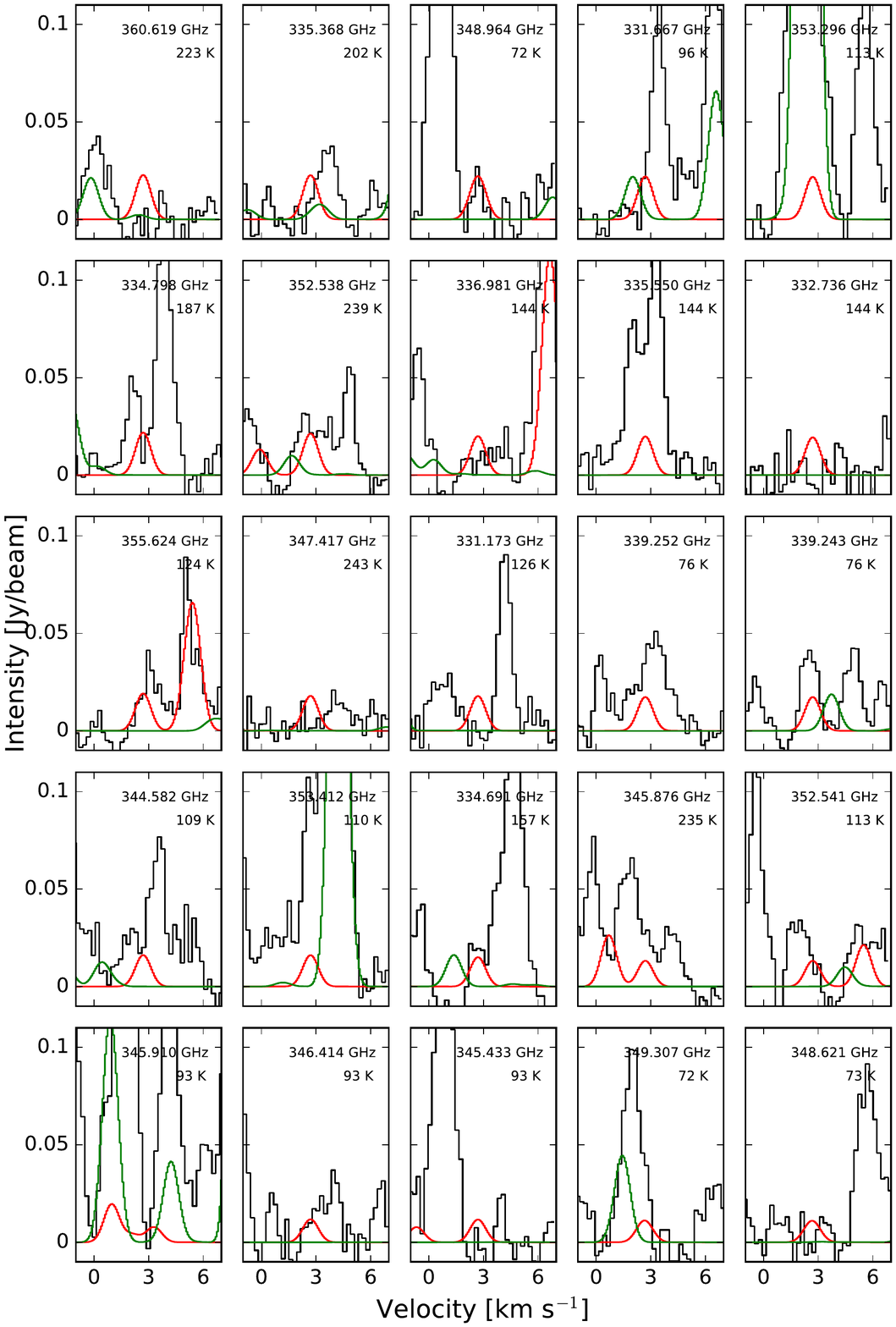}
\caption{Acetone (\acetone{}): Synthetic spectrum in red and reference model in green superimposed onto observed spectrum.}
\label{fig:app_ac_12}
\end{figure*}

\clearpage

\nopagebreak
\onecolumn
\section{Detected lines}
\begin{center}
\begin{longtable}{lcccccc}
\caption[]{\label{tlines}Catalog values for the detected propanal, ethylene oxide and acetone transitions and the integrated line strength of the synthetic spectrum. Due to line contamination, the integrated line strength is given for the FWHM of the Gaussian function.}\\
  \hline \hline
  \multicolumn{1}{l}{Transition} &
\multicolumn{1}{c}{Frequency} &  
  \multicolumn{1}{c}{$E_\mathrm{up}$} &
  \multicolumn{1}{c}{$\log_{10}$($A_\mathrm{ul})$}&
    \multicolumn{1}{c}{$\tau$}&
  \multicolumn{1}{c}{$\int_\mathrm{FWHM} \mathrm{I} \, \delta \varv $}  &
   \multicolumn{1}{c}{Detection level} \bigstrut[t] \\
 & $[$MHz$]$ & $[$K$]$ & [s$^{-1}$] & & [J beam$^{-1}$ km s$^{-1}$] &  \bigstrut[t] \\
\hline 
 \endfirsthead
 
 \multicolumn{3}{c}{{\tablename\ \thetable{} -- continued from previous page}} \\
  \hline \hline
  Transition &
Frequency &  
  $E_{\rm up}$ &
  $\log_{10}$($A_{\rm ul})$&
  $\tau$ &
  $\int_\mathrm{FWHM} \mathrm{I} \, \delta \varv $ & Detection level  \bigstrut[t] \\
 & $[$MHz$]$ & $[$K$]$ & [s$^{-1}$] & & [J beam$^{-1}$ km s$^{-1}$] & \bigstrut[t] \\
\hline 
\endhead 

\hline \multicolumn{5}{c}{{Continued on next page}} \\ \hline
\endfoot

\hline
\hline
\endlastfoot
 \multicolumn{7}{ c }{Ethylene oxide (\ethyleneoxide{})} \bigstrut[t] \\
\hline
$ 10 _{  2 ,  8 } -   9 _{  3 ,  7 }$ & 329744.34 &  95 & -3.3018 & 5.33e-01 &  0.248  & 69 \\
$ 10 _{  3 ,  8 } -   9 _{  2 ,  7 }$ & 329748.33 &  95 & -3.3019 & 3.20e-01 &  0.147  & 41 \\
$ 22 _{  5 , 17 } -  22 _{  4 , 18 }$ & 331651.32 &  441 & -3.7416 & 2.58e-02 &  0.064  & 18 \\
$ 22 _{  6 , 17 } -  22 _{  5 , 18 }$ & 331651.32 &  441 & -3.7416 & 1.55e-02 &  0.064  & 18 \\
$ 21 _{  4 , 17 } -  21 _{  3 , 18 }$ & 332345.92 &  392 & -3.8096 & 1.87e-02 &  0.034  & 10 \\
$ 21 _{  5 , 17 } -  21 _{  4 , 18 }$ & 332345.92 &  392 & -3.8097 & 3.12e-02 &  0.034  & 10 \\
\end{longtable}
\end{center}

\newpage

\section{Acetone lines with high K$_{\rm a}$ and low K$_{\rm c}$ quantum numbers}

\begin{table*}
 \caption[]{\label{missing_lines} Acetone lines with high K$_{\rm a}$ and low K$_{\rm c}$ quantum numbers}
 \centering
\begin{tabular}{lc}
 \hline \hline
  Transition & Predicted frequency [MHz] \\
  \hline
    $19_{13, 7}$  -  $18_{12, 6}$ EE & 334948.01    \\
     $18_{13,5}$ -   $17_{12, 6}$ EE & 336133.79 \\
   $18_{14, 5}$ -  $17_{13, 4}$  AE &  338242.61 \\
   $18_{14, 4}$ -  $17_{13, 4}$ EA & 338625.87 \\
     $18_{14, 4}$ -  $17_{13, 5}$ AA & 339459.74 \\
           $18_{12, 6}$  -   $17_{11, 7}$ EE  & 342896.39    \\
      $18_{15, 3}$ -  $17_{14, 3} $    EA  & 344793.57\\
            $19_{14, 6}$  - $18_{13, 5} $ EE & 350205.27   \\
      $19_{14, 6}$  - $18_{13, 5} $    AA & 351215.21 \\
       $19_{14, 6}$  -  $18_{13, 6}$ EE  & 353155.67   \\
       $19_{14, 5}$  -  $18_{13, 6}$ AA & 354673.02 \\
             $19_{14, 5}$  -   $18_{13, 6} $ EE  & 355248.54   \\
      $19_{13, 6}$  - $18_{12, 7}$   AA  & 356644.96\\
      $19_{13,6}$ -  $18_{12, 7}$ EE & 356675.20 \\
              $19_{15, 4}$  -   $18_{14, 5}$  AE & 359197.90   \\
       $19_{15, 5}$  - $18_{14, 4}$ AA    & 359301.10  \\
       $20_{14,7}$  -  $19_{13,6}$ EE & 359843.80 \\
      $20_{14, 7}$   -  $19_{13, 6}$ AA & 360338.51     \\
\hline
\end{tabular}
\tablefoot{These acetone lines appear missing or shifted when comparing the synthetic spectrum with the observed spectrum. This could be due to perturbations by interactions between the (high K$_{\rm a}$, low K$_{\rm c}$) levels and the levels from the lowest torsional excited states \citep{Groner2002}.}
\end{table*}

\end{appendix}

\end{document}